\documentclass[letterpaper,twocolumn,10pt]{article}
\usepackage{usenix2019_v3}
\usepackage[table]{xcolor}
\usepackage[capitalize,nameinlink,noabbrev]{cleveref}
\usepackage{textcomp}

\usepackage[numbers,sort&compress]{natbib}

\usepackage{kotex}

\usepackage{paralist}

\usepackage[table,xcdraw,pdftex,dvipsnames]{xcolor}

\usepackage{algorithm}
\usepackage{algorithmicx}
\usepackage[noend]{algpseudocode}

\usepackage{siunitx}

\usepackage{multirow}
\usepackage{tabularx}
\usepackage{array}
\usepackage{makecell}
\usepackage{booktabs}
\usepackage{threeparttable}

\usepackage[framemethod=tikz]{mdframed}
\usetikzlibrary{shadows}
\usepackage{graphics}
\usepackage{graphicx} 
\usepackage{tikz}
\usepackage{subcaption}

\usepackage[normalem]{ulem}
\usepackage{xspace}
\usepackage{listings}
\usepackage{enumitem}
\usepackage{url}
\usepackage{float}
\usepackage{tcolorbox}
\usepackage{todonotes}
\usepackage{xargs}
\usepackage{pifont}
\usepackage{hyperref}
\usepackage{tikz}
\usepackage[draft, commandnameprefix=ifneeded, commentmarkup=uwave]{changes}

\setdeletedmarkup{\textcolor{red}{#1}}

\usepackage{flushend}  

\usepackage{breakurl}

\definecolor{lightlightgray}{gray}{0.85}

\newmdenv[
    tikzsetting= {fill=blueish!15},
    skipabove=0.33em,
    skipbelow=0.33em,
    linewidth=1pt,
    innerleftmargin=4pt,
    innerrightmargin=4pt,
    innertopmargin=2pt,
    innerbottommargin=2pt,
    linecolor=gray95,
    roundcorner=2pt, 
    shadow=true,
    shadowsize=4pt,
    shadowcolor=black
]{myshadowbox}

\newcommand{\resultbox}[2][]{%
  \tcbset{
    colback=black!5, colframe=black!50,
    boxrule=0.5mm, arc=1mm, boxsep=2mm, left=0.5mm, right=0.5mm, top=0.5mm, bottom=0.5mm
  }
  \begin{tcolorbox}
  \ifthenelse{\equal{#1}{}}
    {}
    {\textbf{\textbf{#1:}} }
  #2
  \end{tcolorbox}
}

\newcolumntype{s}{>{\centering \arraybackslash \hsize=.5\hsize}X}  %

\algnewcommand{\LineComment}[1]{\State \(\triangleright\) #1}

\definecolor{my-full-blue}{HTML}{1F77B4}
\definecolor{my-full-orange}{HTML}{FF7F0E}
\definecolor{my-full-green}{HTML}{2CA02C}
\definecolor{my-full-red}{HTML}{d62728}
\definecolor{my-full-purple}{HTML}{9467bd}

\colorlet{my-blue}{my-full-blue!30}
\colorlet{my-orange}{my-full-orange!30}
\colorlet{my-green}{my-full-green!30}
\colorlet{my-red}{my-full-red!30}
\colorlet{my-purple}{my-full-purple!30}

\definecolor{verylightgray}{rgb}{.99,.99,.99}

\lstdefinestyle{CStyle}{
    language=C,
    backgroundcolor=\color{verylightgray},
    basicstyle=\ttfamily\scriptsize,
    keywordstyle=\color{blue}\bfseries,
    commentstyle=\color{green!50!black},
    stringstyle=\color{red},
    numbers=left,                      
    numberstyle=\scriptsize\color{gray},      
    stepnumber=1,                       
    numbersep=10pt,                     
    tabsize=4,
    showspaces=false,
    showstringspaces=false,
    breaklines=true,
    breakatwhitespace=true,
    captionpos=b,                       
    escapeinside={\%*}{*)},              
    morekeywords={memcpy}
}

\lstdefinestyle{PlainText}{
    basicstyle=\ttfamily\scriptsize,    
    breaklines=true,                      
    breakatwhitespace=true,               
    numbers=none,                         
    showspaces=false,                     
    showstringspaces=false,               
    frame=none,                           
    backgroundcolor=\color{white},        
    commentstyle=\color{black},           
    keywordstyle=\color{black},           
    stringstyle=\color{black},            
}

\lstdefinestyle{diff}{
    backgroundcolor=\color{verylightgray},
    basicstyle=\ttfamily\scriptsize,
    morecomment=[f][\color{red}]{-},      
    morecomment=[f][\color{green!60!black}]{+}, 
    morecomment=[f][\color{gray}]{@@},    
    morecomment=[f][\color{blue}]{diff},  
    morecomment=[f][\color{blue}]{index}, 
    morecomment=[f][\color{blue}]{---},   
    morecomment=[f][\color{blue}]{+++},   
    breaklines=true                       
}

\lstdefinestyle{compactdiff}{
    backgroundcolor=\color{verylightgray},
    basicstyle=\ttfamily\fontsize{7pt}{7pt}\selectfont,
    morecomment=[f][\color{red}]{-},      
    morecomment=[f][\color{green!60!black}]{+}, 
    morecomment=[f][\color{gray}]{@@},    
    morecomment=[f][\color{blue}]{diff},  
    morecomment=[f][\color{blue}]{index}, 
    morecomment=[f][\color{blue}]{---},   
    morecomment=[f][\color{blue}]{+++},   
    lineskip=-0.5pt,
    breaklines=true                       
}

\lstdefinestyle{sanitizer}{
    backgroundcolor=\color{verylightgray},
    basicstyle=\fontsize{6.5pt}{7.8pt}\ttfamily\selectfont,
    morecomment=[l][\color{red}]{==},
    morecomment=[l][\color{red}]{SUMMARY},
    breaklines=true,
}

\renewcommand{\paragraph}[1]{\vspace{5pt}\noindent{\bf #1}\hspace{7pt}}

\usepackage{kotex}  

\definecolor{hongblue}{RGB}{0,100,200}

\usepackage{subcaption} 
\usepackage{placeins}
\usepackage{array}
\usepackage{tabularx}
\usepackage{threeparttable}
\usepackage{float}
\usepackage{cuted}
\begin{document}

\date{}

\captionsetup[sub]{font=tiny}

\newcommand{\NAME}{\textsc{San2Patch}}
\newcommand{\DATA}{\textsc{San2Vuln}}
\renewcommand{\thefootnote}{\fnsymbol{footnote}}

\newcolumntype{C}{>{\centering\arraybackslash}X}

\title{\Large \bf Cross‑National Information Attacks: \\A Two‑Decade Analysis of Troll Behavior in Korea}


\author{
\begin{tabular}{@{}c@{}}
{\normalfont
Jaehong Kim\textsuperscript{1,2,*} \quad
Hyeonseung Kim\textsuperscript{1,*} \quad
Jiseon Kim\textsuperscript{1,2}} %
\\
{\normalfont Alice Oh\textsuperscript{1} \quad
Thorsten Holz\textsuperscript{2} \quad
Wonjae Lee\textsuperscript{1,\textdagger} \quad
Meeyoung Cha\textsuperscript{2,1,\textdagger}} \\ 
\small 
\textsuperscript{1}\textit{Korea Advanced Institute of Science and Technology (KAIST), Daejeon, South Korea} \\
\small 
\textsuperscript{2}\textit{Max Planck Institute for Security and Privacy (MPI-SP), Bochum, Germany} 
\end{tabular}
}

\maketitle

\renewcommand{\thefootnote}{\fnsymbol{footnote}}
\footnotetext[1]{Equal contribution.}
\footnotetext[2]{Corresponding authors.}
\renewcommand{\thefootnote}{\arabic{footnote}}
\setcounter{footnote}{0}

\begin{abstract}
\noindent
Coordinated foreign influence operations pose a growing threat to online platforms, but detecting state-linked troll activity and tracking its evolution remain challenging. 
This paper presents an explainable machine learning framework for theory-guided detection and longitudinal analysis of suspected trolling within Korean online news comment sections.
Our hierarchical model classifies comments along three dimensions central to influence campaigns: foreign origin, moral-emotional framing, and target country. 
To support explainability, it also extracts brief span-level textual evidence that provides human-interpretable rationales. We apply the approach to 112M South Korean news comments authored by 4M users over nearly 20 years, identifying 23,998 accounts exhibiting behavior consistent with coordinated manipulation. 
Analyzing these accounts, we find that they predominantly rely on morally condemning rhetoric rather than direct promotion of foreign-aligned narratives; this rhetoric receives significantly higher user engagement.
Among the highest-engagement comments, the moral condemnation most frequently targets domestic political figures (e.g., presidents or party leaders) on both the left and the right, potentially amplifying polarization.
Our framework supports transparent platform governance through explainable, evidence-based moderation. These observed rhetorical and engagement patterns can inform how platforms and observatories prioritize defenses and intervene before harmful narrative-target combinations achieve widespread reach.

\end{abstract}

\section{Introduction}
\label{sec:introduction}

\begin{figure*}[t]
    \centering
    \includegraphics[width=.95\textwidth]{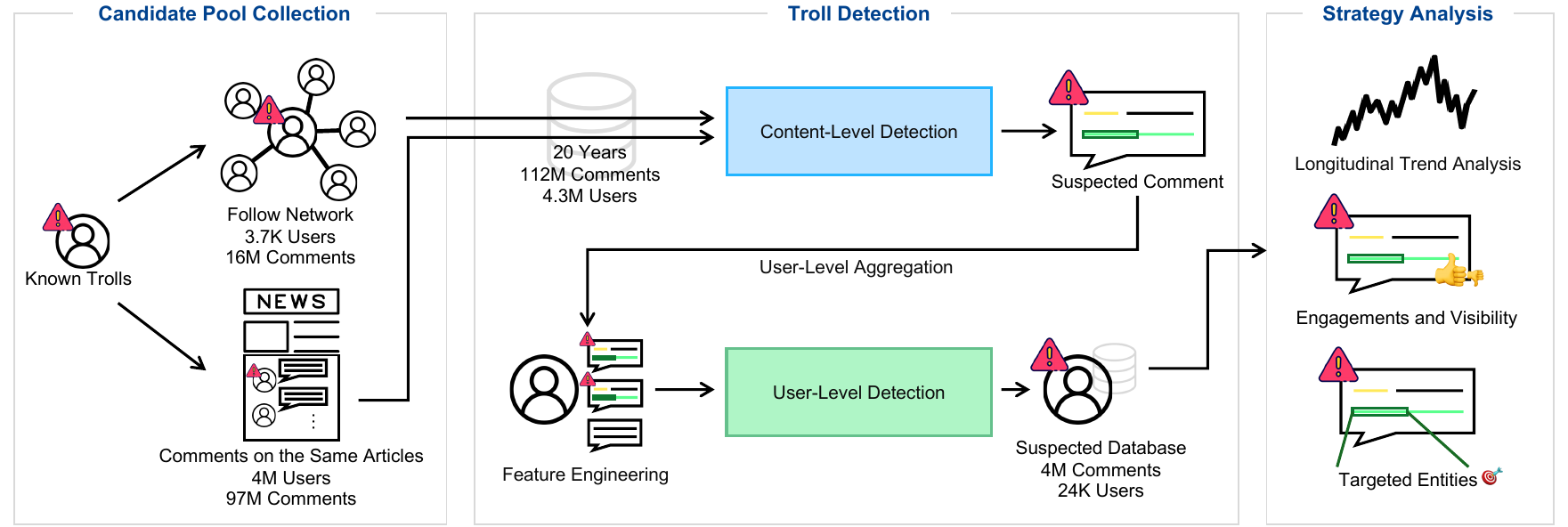}
    \caption{Framework for analyzing suspected troll strategies, where explainable AI connects detection to subsequent strategy analysis.
    (i) Candidate pool collection (\Cref{sec:candidate_pool}): expand from known trolls via follow networks and shared-article commenting.
    (ii) Troll detection (\Cref{sec:Explainable,sec:userlevel_detection}): run an explainable content detector, then aggregate outputs for user-level detection.
    (iii) Strategy analysis (\Cref{sec:strategy_analysis}): analyze temporal trends, engagement patterns, and targeted entities. }
    \label{fig:overview}
    \vskip -0.1cm
\end{figure*}

As information operations increasingly unfold online, adversaries can scale campaigns that target public cognition rather than physical territory, amplifying coordinated narratives to large audiences~\cite{bernal2020cognitive, claverie2022cognitive}.
Prior research has documented troll campaigns that disrupt elections, distort public health discourse, and exploit politically sensitive issues to intensify polarization across multiple democracies~\cite{alizadeh2020content,broniatowski2018weaponized,arif2018acting, kelton2019australia,hung2022china}. 
From a security perspective, such operations threaten the integrity of democratic debate and motivate defenses that enable reliable detection and auditable analysis over time~\cite{starbird2019disinformation, schroeder2026malicious,mirza2023tactics}.

Despite progress in machine learning methods for detection~\cite{saeed2022trollmagnifier, xiao2025sokML,hanley2024specioussites}, two gaps still hinder effective defense against influence operations. 
First, many systems produce binary or coarse labels with limited evidence explaining why content or accounts are flagged, restricting auditability and operational use. Enforcement actions such as warnings, down-ranking, or account restrictions are frequently contested by users and difficult for moderators to justify without an evidence-based rationale~\cite{myers2018censored, jhaver2019does}. Second, we lack systematic evidence on how rhetorical strategies achieve platform visibility and thus reach large audiences. While prior work describes organizational and operational structures~\cite{recabarren2023strategies}, it remains unclear which \emph{content-level tactics} reliably surface through ranking mechanisms at scale. Theory suggests that `condemning' target countries may be more effective than overt propaganda~\cite{hung2022china}, but large-scale, longitudinal validation remains scarce.

This paper presents a three-stage framework that connects explainable troll detection to longitudinal analysis of influence strategies, enabling both large‑scale deployment and systematic auditing. Figure~\ref{fig:overview} depicts an overview of our approach.
The framework is evaluated on a newly assembled corpus of 112 million comments spanning nearly two decades from the most widely used news portal in South Korea, providing a uniquely long observational window into the evolution of foreign‑influence rhetoric.

At a high level, we propose the following three steps: (1)
\textbf{Candidate pool collection:} Starting from a verified seed set of 70 troll accounts publicly released by the Institute for National Security Strategy in Korea~\cite{INSS2024ForeignInfluence}, we expand to a candidate pool via two complementary traces: users connected through the follow network and users who commented on the same articles as the seeds. This process leads to 112 million comments spanning 2006--2025, suitable for detecting coordinated influence beyond the initially identified accounts. 
(2) \textbf{Troll detection:} We run an explainable content-level detector that predicts hierarchical labels grounded in cognitive warfare research and social-psychological theory, and outputs span-level rationales as textual evidence. We fine-tune an LLM for this supervision and distill it into a lightweight model with roughly 0.1 billion parameters, enabling cost-effective inference over the full corpus.
We then aggregate comment-level outputs into user-level features to identify \emph{suspected troll accounts}. 
(3) \textbf{Strategy analysis:} We quantify how rhetorical strategies of these accounts evolve over time, how they translate into engagement and visibility under the platform's ranking dynamics, and which entities are repeatedly targeted. This analysis helps identify  which tactics surface and persist in public discourse and provides an empirical basis for defensive prioritization.

Across five presidential administrations in South Korea during the measurement period, one pattern emerges gradually yet persistently: suspected troll accounts increasingly post comments expressing the \emph{Condemning Korea} narrative, which receive higher user engagement within the target country than direct promotion of foreign agendas. We also find that the influx of troll-like users increases around election periods and during major diplomatic tensions, indicating responsiveness to political context. 
In fact, \emph{Condemning Korea} comments are more likely than \emph{praising} comments to surface in top-ranked threads. Evidence further indicates that political leaders from both right-leaning and left-leaning camps are repeatedly targeted by moral condemnation, suggesting an emphasis on intensifying polarization rather than promoting a single faction.
Beyond detection, our framework provides actionable evidence for mitigation and response. 
Hierarchical predictions with span-level rationales can support evidence-based moderation workflows, and our analysis offers empirical guidance for prioritizing defensive attention toward high-visibility rhetoric. 
We discuss these implications for platform systems and defensive prioritization.

\smallskip 
We make the following contributions.
\begin{compactitem}
    \item[1.] We release nearly a \textbf{two-decade dataset} of 112 million South Korean news comments authored by 4 million distinct users.\footnote{Due to the nature of this research, examples contain biased content.} By linking publicly identified troll accounts to their interaction contexts, the dataset enables large-scale longitudinal analysis of influence activity.
    
    \item[2.] We design an \textbf{explainable troll detector} that produces hierarchical labels and span-level rationales. Grounded in cognitive warfare and social-psychological theory, the model uses knowledge distillation for fast inference over 112 million comments, enabling auditability and supporting transparent analysis beyond black-box detection.

    \item[3.] 
    Our data reveal \textbf{influence strategies}, including sharp activity spikes around elections and a growing preference for condemning rhetoric over direct promotion of foreign-aligned narratives. We show that condemning comments targeting local politicians are disproportionately likely to appear in top-ranked threads, providing actionable guidance for defensive prioritization.
\end{compactitem}

\smallskip 
These contributions advance our understanding of foreign influence within online news comment sections. To support future research, we share our dataset of news comments, verified seeds, and model-detected troll accounts (detailed in the Open Science section).\footnote{\url{https://doi.org/10.5281/zenodo.20257085}} This repository provides the empirical foundation for important future directions, such as formalizing adversarial threat models that account for multi-faceted political agendas and the strategic deployment of moral-emotional rhetoric.


\section{Foreign Influence Dataset} 
\label{sec:candidate_pool}

\subsection{Naver News Platform}
This is South Korea's dominant news platform, which is the primary news source used by 63\% of the Korean adult population~\cite{newman2025digital, kpf2025media}. Similar to Google News, it operates a platform-centric ecosystem where major media outlets publish articles directly on the site. Users can post comments beneath each article, like or dislike other users' comments, and sort comments by vote-based ranking.

\paragraph{Comment Ranking and Visibility.}
Visibility is determined by a like ratio, calculated as likes divided by total engagement (likes plus dislikes). This mechanism elevates comments that attract broad agreement while reducing the visibility of comments that provoke strong backlash or controversy. Because top-ranked comments receive disproportionate exposure, the highest positions in a thread offer an efficient channel for broadcasting narratives~\cite{jeong2020identifying}. Since readers rely on these comments as heuristic cues for the prevailing opinion climate~\cite{han2025commenters}, the comment section has become a high-stakes arena for influence operations seeking to manipulate public discourse~\cite{INSS2024ForeignInfluence}.

\paragraph{Follow Network.}
Naver introduced a follower-following feature that allows users to subscribe to specific comment authors and view their activity. Followed comments receive priority placement on article pages (up to 100 comments, shown in reverse chronological order) and are aggregated in a dedicated "Followed Comments" feed. The follow relationship is asymmetric: if a user blocks another, the blocked user cannot follow, reply to, or upvote their comments. 

\subsection{Data Collection}
The Institute for National Security Strategy in Korea identified and released in 2024 a set of 70 foreign-actor accounts that persistently promoted narratives aligned with specific foreign interests within Naver News comment sections~\cite{INSS2024ForeignInfluence}. We treat these accounts as \textit{known trolls} and use them as seeds for our data collection strategy. In contrast to this `gold set,' our findings are classified as `suspected' state-linked activity, acknowledging that these results are model-generated predictions rather than manually verified instances.

\paragraph{Troll Activity.}
We first collected all comments posted by the 70 known trolls, yielding 356,378 comments across 262,875 articles with an average of 5,091 comments per account. This unusually high activity rate indicates sustained influence efforts and suggests the use of automation.

\paragraph{Candidate Pool Collection.}
\label{Candidate_Pool_Construction}
To detect additional troll accounts, we constructed a large candidate pool using two complementary approaches.
First, we retrieved the complete follower and following lists of all known trolls and collected comments written by users directly connected to them, yielding 16,085,636 comments from 3,703 users. Second, we collected all comments posted under the 262,875 articles where known trolls had been active, producing 97,765,429 comments from 4,046,948 unique users. This dual approach captures both the social network surrounding known trolls and the broader user population engaged with troll-targeted content. 
After merging these datasets with the known troll corpus and removing duplicates, we obtained 112,658,554 comments authored by 4,047,831 users as our target dataset for troll detection, covering the period from April 2006 to March 2025. Summary statistics are in Appendix Table \ref{tab:troll_user_dataset}.

\section{Explainable Content-Level Troll Detection
}
\label{sec:Explainable}

\begin{figure}
    \centering
 
    \makebox[\columnwidth][c]{\includegraphics[width=1.05\columnwidth]{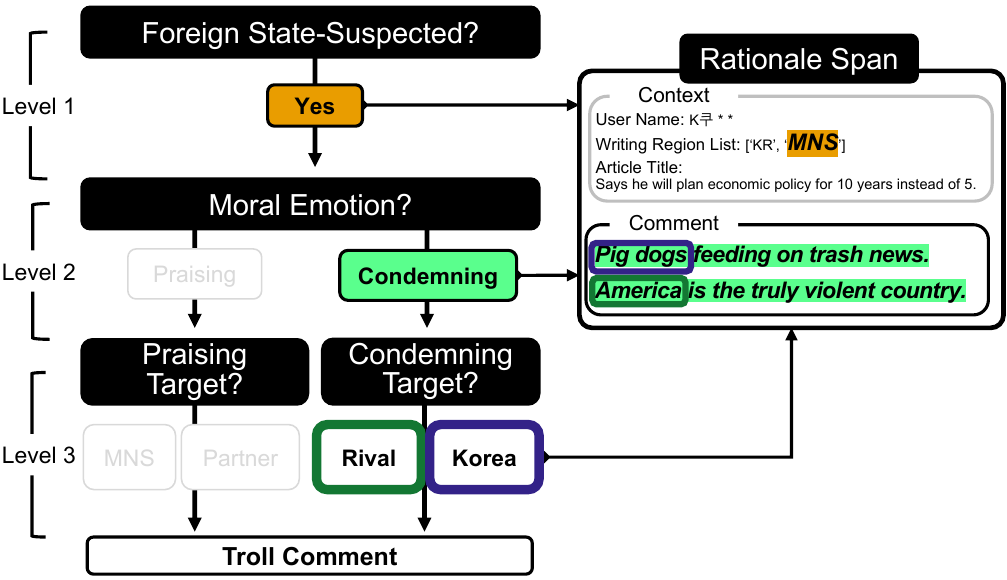}}
    \caption{A framework for explainable comment-level troll detection. It performs multi-level classification (foreign state suspicion, moral emotion, and target country) while extracting level-specific rationales to justify the final troll prediction. MNS denotes a \emph{major neighboring state} of South Korea.}
    \label{fig:framework}
\end{figure}

We present an explainable troll detection model that combines a hierarchical labeling scheme with a scalable knowledge distillation pipeline. We begin by conducting human annotation, defining key influence-messaging cues (such as foreign state origin, moral emotions, and targeted countries) and annotating comments with short span-level rationales to create gold-standard data. We then fine-tune GPT-4.1 on these gold labels to generate hierarchical predictions with supporting spans and apply it to annotate 50,000 candidate comments drawn from our 112 million comment corpus. Finally, we apply knowledge distillation by training lightweight open source models on the GPT-annotated data, enabling fast and cost-efficient inference while preserving accurate hierarchical classifications and span-based explanations.

\subsection{Data Categorization}

\paragraph{Design Rationale}\label{subsec:rationale-hierarchical} We designed a three-level hierarchical labeling framework to operationalize the detection of foreign influence operations on Naver News, grounded in research on cognitive warfare and social psychology. Inspired by explainable AI methods developed for hate speech detection~\cite{zampieri-etal-2019-predicting, jeong-etal-2022-kold}, our framework jointly assigns class labels and their supporting span-level rationales to ensure the explainability of the classification process. Figure~\ref{fig:framework} provides an overview of the annotation process.

\paragraph{Level 1: Foreign state-suspected origin.} 
According to NATO, cognitive warfare involves the weaponization of public opinion by external actors to influence policy and destabilize public institutions~\cite{bernal2020cognitive}. This definition implies that the attacking entity must originate outside the target country. The first level therefore identifies whether the comment author is likely of foreign state-suspected origin, based on linguistic, cultural, and behavioral markers including writing region metadata, username patterns, grammatical patterns, and culturally-specific expressions.

\paragraph{Level 2: Moral emotions.} 
The second level identifies moral emotions that have been empirically linked to online polarization~\cite{brady2017emotion,van2024social}. Prior research shows that trolls aim to intensify division and emotional conflict~\cite{simchon2022troll,arif2018acting,broniatowski2018weaponized}. According to Brady et al.~\cite{brady2020mad}, moral emotions such as \textit{other-condemning} (anger, disgust, contempt) and \textit{other-praising} (admiration, elevation) play central roles in this process. \textit{Other-condemning} rhetoric establishes in-group versus out-group boundaries by attacking others' moral integrity, while \textit{other-praising} rhetoric reinforces in-group cohesion by elevating the moral status of one's own side. Empirical studies in Korea show that these emotions are strong predictors of online political polarization~\cite{kim2024moral}, and research on foreign propaganda strategies indicates that cognitive-warfare campaigns often rely on these two rhetorical pathways~\cite{hung2022china}.

\paragraph{Level 3: Target country.} 
The third level identifies the country toward which the moral emotion is directed. Although mapping these expressions to a specific nation can be challenging due to text ambiguity (requiring an "Unknown" category), this classification is essential for tracking foreign influence. Pinpointing the national target enables a fine-grained analysis of which states are condemned or praised, revealing the strategic focus of the underlying campaigns. Following established geopolitical classifications in East Asia~\cite{csis2023tenfortaiwan}, we define five target categories, assigned whenever a comment directs moral emotion toward a country's leaders, citizens, or sociopolitical issues.
\begin{itemize}
    \item \textbf{Trolled region}: South Korea
    

    \item \textbf{MNS (Major Neighboring State)} \textbf{[Anonymized]}: suspected troll origin of foreign state-linked actors ($\star$The country label is anonymized to keep the analysis focused on technical patterns rather than geopolitical conflict.)
    

    \item \textbf{Partner}: states aligned with the origin like Russia
    \item \textbf{Rival}: states in strategic opposition like the US or Japan
    \item \textbf{Unknown}: no specific country can be identified
\end{itemize}
{
\begin{table}[t]
\centering
\caption{Inter-annotator agreement across hierarchy.}
\label{tab:iaa_scores}
\begin{tabular}{lc}
\toprule
\textbf{Labeling Level} & \textbf{Krippendorff's $\alpha$} \\
\midrule
Foreign state-suspected & 0.994 \\
Moral emotion & 0.512 \\
Condemning target country & 0.625 \\
Praising target country & 0.706 \\
Troll comment label & 0.813 \\
\bottomrule
\end{tabular}
\end{table}
}

\noindent
\textbf{Identifying troll content.}
We use these three levels hierarchically as operational criteria for identifying troll-like influence messaging. A comment is labeled as \textit{suspected troll content} when all three conditions are met: (1) the commenter is located in foreign state-suspected, as inferred from an IP-based region marker and non-native linguistic features (Level 1); (2) the comment expresses a moral emotion (Level 2); and (3) the identified target aligns with the actor's ideological stance (Level 3).  In practice, \textit{condemning} rhetoric directed at South Korea or rivals is labeled as troll content, whereas \textit{praising} rhetoric directed at a major neighboring state or its partner is likewise labeled as troll content.

{
\begin{table*}[t]
\centering
\caption{Performance comparison of fine-tuned GPT and lightweight models on class- and span-level tasks (Macro F1). Bold indicates the best in each column.}
\label{tab:unified_performance_comparison}
\small
\renewcommand{\arraystretch}{1.15}

\newcolumntype{P}[1]{>{\centering\arraybackslash}p{#1}}

\resizebox{0.95\textwidth}{!}{
\begin{tabular}{
l 
| P{1.1cm} P{1.1cm}      
| P{1.1cm} P{1.1cm}      
| P{1.1cm} P{1.1cm}      
| P{1.1cm} P{1.1cm}      
| P{1.1cm}                
| P{1.3cm}                
}
\toprule
\multirow{2}{*}{\textbf{Model / Shot}} & 
\multicolumn{2}{c|}{\makecell{\textbf{Foreign state-}\\\textbf{suspected}}} & 
\multicolumn{2}{c|}{\textbf{Moral emotion}} & 
\multicolumn{2}{c|}{\textbf{Condemning target}} & 
\multicolumn{2}{c|}{\textbf{Praising target}} & 
\textbf{Troll} & 
\multirow{2}{*}{\makecell{\textbf{Average}\\\textbf{F1}}} \\
\cmidrule(lr){2-3}\cmidrule(lr){4-5}\cmidrule(lr){6-7}\cmidrule(lr){8-9}\cmidrule(lr){10-10}
& \textbf{Class} & \textbf{Span} 
& \textbf{Class} & \textbf{Span} 
& \textbf{Class} & \textbf{Span} 
& \textbf{Class} & \textbf{Span} 
& \textbf{Class} & \\
\midrule
  GPT (100)        & 0.9940 & 0.8388 & 0.7749 & 0.7740 & 0.5855 & 0.4510 & \textbf{0.6931} & 0.4019 & 0.9468 & 0.7178 \\                           
  GPT (200)        & 0.9950 & 0.9089 & 0.7805 & 0.7810 & 0.4998 & 0.4712 & 0.4747 & 0.4296 & \textbf{0.9508} & 0.6991 \\                           
  GPT (300)        & 0.9950 & 0.9466 & 0.7776 & 0.7890 & 0.5174 & 0.4878 & 0.5930 & \textbf{0.4309} & 0.9478 & 0.7206 \\                           
  GPT (400)        & 0.9960 & \textbf{0.9767} & 0.7715 & \textbf{0.7902} & 0.5374 & \textbf{0.4991} & 0.5202 & 0.3914 & 0.9458 & 0.7143 \\
  \midrule
  KcBERT             & \textbf{0.9970} & 0.8075 & 0.8459 & 0.5958 & 0.6891 & 0.4026 & 0.6160 & 0.3934 & 0.9190 & 0.6963 \\
  KLUE-RoBERTa          & 0.9960 & 0.6308 & 0.8544 & 0.5636 & 0.6505 & 0.3907 & 0.6093 & 0.3384 & 0.9170 & 0.6612 \\
  \textbf{KcELECTRA} & 0.9960 & 0.8175 & \textbf{0.8806} & 0.6304 & \textbf{0.7119} & 0.4417 & 0.6892 & 0.4042 & 0.9279 & \textbf{0.7222} \\
\bottomrule
\end{tabular}}
\end{table*}
}

\subsection{Annotation}

\paragraph{Human Labels}
We compiled a sample of 1,500 instances from the known troll corpus and the candidate pool for human annotation. This included 825 comments from known trolls to ensure coverage of their characteristic messaging patterns, and 675 comments from  users who commented on the same articles. This comparative design captures accounts engaging with identical content and topics but potentially exhibiting distinct stances or behaviors. To mitigate emotional bias, an off-the-shelf moral emotion classifier~\cite{kim2024moral} was used to ensure that at least 100 comments per group contained \textit{other-condemning} and \textit{other-praising} expressions. The remaining samples were randomly selected.

Each comment was presented alongside its metadata, including the username, writing region, and the corresponding news article title. Annotation was conducted using the Label Studio platform~\cite{labelstudio} by three authors. Annotators assessed three hierarchical levels for each comment, highlighting specific text spans supporting their judgments at each level. Within this framework, the foreign state-suspected designation was treated as binary, whereas moral emotions and target countries were evaluated as multi-label tasks.

Only samples that achieved majority agreement across annotators for both class labels and span rationales were retained, yielding a gold label dataset of 1,452 comments. We tested annotation reliability using Krippendorff's $\alpha$.  Table~\ref{tab:iaa_scores} shows near-perfect agreement for \textit{foreign state-suspected} ($\alpha=0.994$), as metadata such as writing regions provided explicit cues. Agreement for \textit{moral emotion} was moderate ($\alpha=0.512$) and the lowest across our categories due to its subjective nature, although it remains comparable to or higher than existing emotion-labeling benchmarks~\cite{field2022analysis, demszky2020goemotions}. The final \textit{troll} comment label achieved substantial agreement ($\alpha=0.813$).

\paragraph{GPT Labels}
We divided the 1,452 annotated samples into 400, 52, and 1,000 instances for training, validation, and testing. Unlike in-context learning—which relies on a model's prior knowledge and a few illustrative examples—we fine-tuned the model to update its internal parameters based on human-annotated knowledge~\cite{brown2020language}. Fine-tuning was conducted using OpenAI's \texttt{gpt-4.1-2025-04-14} model, the most advanced fine-tuning API available at the time of experimentation (August 2025)~\cite{achiam2023gpt}.  We trained models with subsets of 100, 200, 300, and 400 samples. The best performance was achieved with 300 samples, resulting in the highest overall F1-score on classification and span-labeling tasks (Average F1 = 0.7206). Detailed results are presented in Table~\ref{tab:unified_performance_comparison}.

To scale annotation beyond the human-labeled samples, we employed a two-stage pipeline. First, we used a pilot ELECTRA-based binary classifier~\cite{lee2021kcelectra} trained on the same 1,452 samples to filter the 112M comment corpus. Because troll comments are relatively rare, we sampled 40,000 comments predicted as troll and 10,000 predicted as non-troll, yielding 50,000 candidates for hierarchical annotation. Second, we applied the fine-tuned GPT model to these candidates to produce hierarchical labels, including both class and span annotations. 
We removed samples where predicted spans did not match or exceeded text boundaries, retaining 49,745 high-quality samples. Minor span offset discrepancies (e.g., one- or two-character shifts), commonly observed in similar studies~\cite{hasanain-etal-2024-large,ding2025span}, were corrected by realigning each span to the closest valid text segment. The distribution of samples across hierarchical classes is provided in Appendix Table~\ref{tab:train_data_distribution}.

\begin{figure*}[t]
\centering
\includegraphics[width =\textwidth,keepaspectratio]{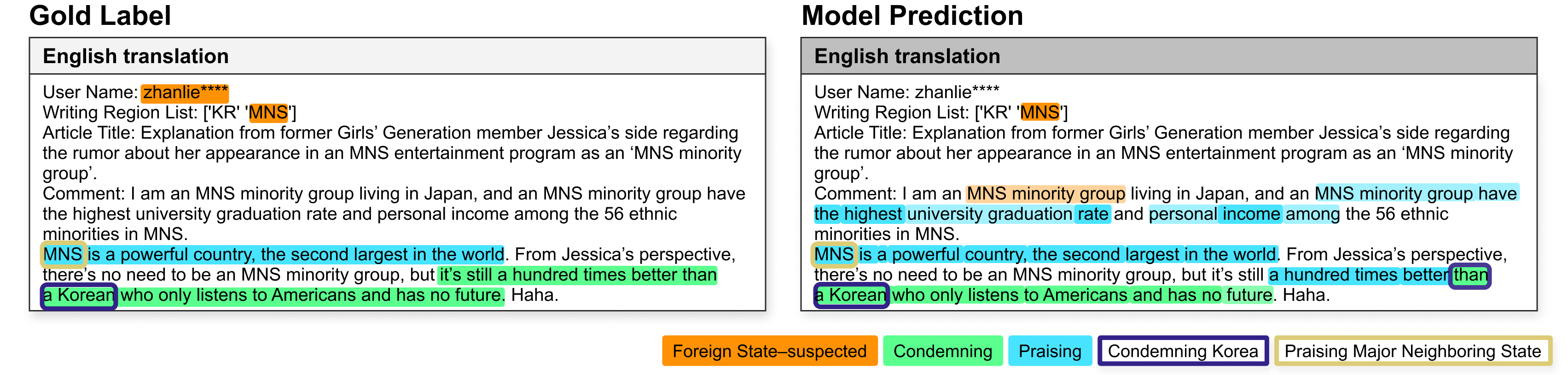}
\caption{
Illustrative output from the final model, with rationale-related spans highlighted. Each span corresponds to a predicted label: \textit{Foreign state-suspected} (filled orange), \textit{Other-condemning} (filled green), \textit{Other-praising} (filled sky blue), \textit{Condemning Korea} (outlined navy), and \textit{Praising MNS} (outlined gold). Span opacity indicates the predicted probability for that label. ``MNS'' refers to \textit{a major neighboring state}. Examples translated from Korean using GPT-5; original Korean outputs in Figure~\ref{fig:rationale_example_kor} in the Appendix.}
\label{fig:rationale_example}
\end{figure*}

\subsection{Knowledge Distillation}
We improve scalability and reduce computational cost by distilling the hierarchical representations learned from the GPT-annotated dataset into lightweight models. We used models pretrained on Korean corpora, including KcBERT~\cite{lee2020kcbert}, KLUE-RoBERTa~\cite{park2021klue}, and KcELECTRA~\cite{lee2021kcelectra}, each with approximately 0.1B parameters. A total of 49,745 GPT-annotated samples were used for model training and validation, with 80\% allocated for training and 20\% for validation.

We built multiple classifiers for each subtask. Binary classification is used to detect final troll content and foreign state-suspected content. Multi-label classification is applied to moral emotions and their corresponding targets, as a single comment can express multiple emotions and reference multiple targets. Span prediction is formulated as a token-level multi-class classification problem by converting spans into BIO tags, where each token is assigned one of the predefined tags. The model consists of 18 classifiers: 2 for binary classification (troll content and foreign state-suspected), 3 for multi-label classification (moral emotions, condemning targets, and praising targets), and 13 for multi-class classification using BIO tagging (1 for a foreign state-suspected span, 2 for moral emotion spans, and 10 for moral emotion target spans). Figure~\ref{fig:rationale_example} provides an illustrative example of the final distilled model's predictions on a human-annotated sample, highlighting token-level rationale spans and their associated labels (with opacity indicating predicted probabilities). The predicted outputs can be compared against the span-level ground truth used for evaluation.

All three models were trained using the same hyperparameters: the AdamW optimizer~\cite{loshchilov2019adam}, a learning rate of 2e-5, weight decay of 1e-2, and 100 epochs, with early stopping applied to prevent overfitting. Binary cross-entropy was used as the loss function for binary and multi-label classification tasks, while cross-entropy loss was used for multi-class classification. Model performance was evaluated using Macro F1. The KcELECTRA-based model achieved the best performance in most categories and was selected as the final model. The evaluation was conducted on 1{,}000 human-annotated test samples, and performance results are shown in Table~\ref{tab:unified_performance_comparison}. Following model selection, inference was performed on 356{,}378 comments from known trolls and 112 million comments from the candidate pool. To assess the model's robustness to surface-level wording changes, we used an LLM to paraphrase the test samples and observed comparable F1 scores, suggesting that the model captures the target concepts rather than relying primarily on surface phrasing. Details are provided in Appendix~\ref{sec:robustness_paraphrase}.



\section{User-Level Troll Detection}
\label{sec:userlevel_detection}

We identify suspected troll users by combining outputs from our explainable content‑level classifier with user‑level behavioral signals informed by prior work~\cite{saeed2022trollmagnifier,kireev2025characterizing,hanley2024specioussites}. We represent each user with a multi-dimensional feature set that captures both substantive messaging and behavioral patterns.

\subsection{Non-Troll Ground Truth}
\label{subsec:nontroll_ground_truth}

While we have a set of 70 known troll accounts, user-level modeling requires a robust ground truth set of non-troll users. To construct this set while controlling for topical context and mitigating class imbalance, we randomly selected 100 candidate users who commented on the same articles as known trolls and sampled 20 comments per user (2,000 comments total) for manual review. Three researchers independently reviewed the sampled comments and made a binary, user-level judgment, using the three-level hierarchy in Section~\ref{subsec:rationale-hierarchical} as coding guidelines (Level~1: foreign state-suspected origin; Level~2: moral emotions; Level~3: target country). 

We applied a highly conservative criterion: an account was designated as a non-troll only if all three annotators unanimously agreed that none of the 20 evaluated comments met our troll classification thresholds. This strict procedure yielded 81 confirmed non-troll users to serve as the negative class for our user-level model.

\begin{figure*}[t]
    \centering
    \includegraphics[width=0.99\textwidth]{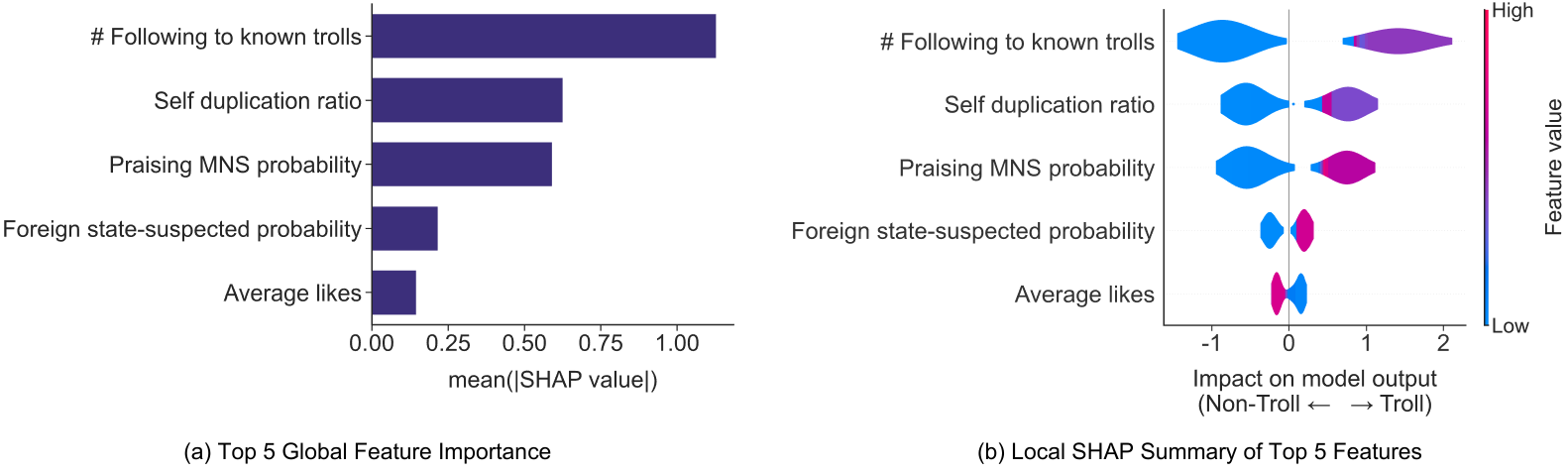}
\caption{Global feature‑importance and SHAP analyses for the user‑level troll‑detection model. \textbf{(a)} The top five features ranked by mean absolute SHAP value, averaged across 10‑fold cross‑validation. \textbf{(b)} SHAP summary plots for the same features, showing the distribution and direction of their effects across users. Higher \textit{frequency of following ties to known trolls} and higher \textit{Praising MNS} probability shift predictions toward the Troll class, whereas higher \textit{average likes} shifts predictions toward Non-trolls. 
}
    \label{fig:combined_ml_results}
    \vskip -0.1cm
\end{figure*}

\subsection{Feature Engineering}

\paragraph{Aggregated Explainable Features.}
For each user, we aggregate the continuous probability outputs from our content-level classifier rather than binarized class labels. For each comment $c$, the classifier produces probabilities for features $f$ aligned with the three hierarchical levels; we then compute the user-level mean across the set of comments authored by user $u$:
\begin{equation}
\bar{p}_u^{(f)} = \frac{1}{|C_u|} \sum_{c \in C_u} p_c^{(f)},
\end{equation}
where $C_u$ denotes the set of comments authored by $u$, $p_c^{(f)}$ is the probability that comment $c$ exhibits feature $f$ (e.g., \textit{Foreign State-Suspected} probability, \textit{Praising MNS} probability), and $\bar{p}_u^{(f)}$ is the user-level mean probability. This aggregation minimizes information loss and captures how strongly and consistently a user exhibits each rhetorical characteristic across their commenting history. We compute 14 features:

\begin{itemize}
    \item Foreign state-suspected origin (Level 1): User-level mean of the classifier's per-comment probabilities for this label.
    
    \item Moral emotions (Level 2): User-level means of the classifier's per-comment probabilities for \textit{other-condemning} and \textit{other-praising}.
    
    \item Target country (Level 3; 10 features): User-level means of per-comment probabilities for \textit{condemning} and \textit{praising} each target category (i.e., \textit{Korea}, \textit{MNS}, \textit{rivals}, \textit{partners}, and \textit{Unknown}).
    
    \item Troll comment probability: User-level mean of the classifier's per-comment troll probability.
\end{itemize}

\paragraph{Behavioral and Metadata Features.}
Following prior methods~\cite{saeed2022trollmagnifier, kireev2025characterizing}, we incorporate 10 behavioral and metadata features commonly associated with influence operations. Together with our content‑based signals, this yields in total 24 user‑level features used for classification, including: 
\begin{itemize}
    \item Exact match overlap: Proportion of a user's comments exactly matching known troll posts.\footnote{To ensure meaningful content matching, we restrict our analysis to comments meeting a minimum length threshold ($\ge$10 characters and $\ge$3 tokens).}
    
    \item Account lifespan: Temporal duration (in hours) between a user's first and last observed comments.
    
    \item Average comment length: Mean character count per comment authored by the user.
    
    \item Engagement statistics (3 features): Mean likes, dislikes, and replies received per comment.
    
    \item Self-duplication ratio: Proportion of repetitive content, defined as $1- (\text{unique comments}/\text{total comments})$. 
    
    \item Following known trolls: Frequency of ties or outdegree to known trolls.
    
    \item Followed by known trolls: Frequency of incoming ties from known trolls. See Appendix~\ref{sec:follow_analysis} for network details.

    \item Comment volume: Total comments per user.
\end{itemize}

\begin{table*}[t]
\centering
\small
\caption{Illustrative examples for each target country category in comments by model-detected trolls, translated using GPT-5.}
\label{tab:stance_examples}
\renewcommand{\arraystretch}{1.5}

\newcolumntype{C}[1]{>{\centering\arraybackslash}m{#1}}

\begin{tabularx}{\textwidth}{C{3.3cm}|X}
\toprule
\multicolumn{1}{c|}{\textbf{Target Country}} & \multicolumn{1}{c}{\textbf{Example}} \\
\midrule

Condemning Korea ~~~~(i.e., Trolled Region) &
A demon race, the Koreans, and a demon state, South Korea. They have maintained a conscription-slavery system for 60 years that serves no purpose other than enslaving their own citizens. These Koreans loudly complain about forced labor from 70 years ago that they never personally experienced, yet remain silent about the ongoing forced labor imposed by their own government on Korean men today.
\\
\midrule

Condemning Rival ~~~~~(i.e., Opposing State)&
The United States is getting desperate over its debt. Are they going to print more dollars again? haha. What about inflation? haha. In 2008, MNS bought U.S. bonds and pulled them through the crisis. So what now? This time MNS will not step in.
\\
\midrule

Praising MNS ~~~ ~~~ (Major Neighboring State)  &
Korea does not seem suited for American-style democracy. It should learn from MNS-style socialism. Why does Korea keep acting up? Within 10 years, MNS will become the world's number one economy. For 5,000 years, MNS was the center of the world, and now it is returning to its rightful place. MNS is a big country, while Korea is a small country. If Korea continues like this, what will happen? The United States is far away, but MNS is close. MNS and Korea should be good friends.
\\
\midrule

~~~ Praising Partner ~~~ $~$ ~~~(i.e., Aligned State) &
That comedy-idiot and his underlings all ran off and hid in civilian residential areas. Don't show mercy to civilians—just grind the whole city into dust. Stay strong, mighty Russia. Putin, you are a hero.
\\

\bottomrule
\end{tabularx}
\end{table*}

\subsection{Evaluation}
We evaluate user-level troll detection using a total of 151 users (70 known trolls and 81 non-troll users). 
For each user, we extract 24 features and train four supervised classifiers: SVM, Random Forest, LightGBM, and XGBoost. 
Using 10-fold cross-validation, the models achieve F1 scores ranging from 0.91 to 0.94: SVM reaches 0.91, Random Forest and LightGBM reach 0.93, and XGBoost performs best with 0.94.

\paragraph{Feature Importance Analysis.} 
To understand the model's decision-making process, we analyze feature importance using SHAP values derived from the XGBoost classifier (Figure~\ref{fig:combined_ml_results}(a)), which quantify each feature's contribution to the model output in a unified, model-agnostic manner\cite{lundberg2017unified}. The five most influential features are the frequency of following ties to known trolls, self duplication ratio, the user-level mean \textit{praising MNS} probability, the user-level mean \textit{Foreign State-Suspected} probability, and average likes.

The local SHAP summary (Figure~\ref{fig:combined_ml_results}(b)) further illustrates how feature values shift predictions toward the Troll or Non-troll class, revealing both the directionality and heterogeneity of feature effects across users. In particular, higher values of content-based features, especially \textit{praising MNS} and \textit{Foreign State-Suspected} probabilities, tend to push predictions toward the troll class. Overall, model-detected trolls exhibit higher values in network and content-based features, whereas higher average likes are more characteristic of non-troll users.

\subsection{Prediction}
We applied the trained XGBoost user-level detector to the merged candidate pool described in \Cref{Candidate_Pool_Construction}, comprising 112M comments authored by 4,047,680 users. 
The candidate pool was constructed by merging (i) the \textit{article-based candidate pool} of users who commented on the same articles as known trolls and (ii) the \textit{follow-network candidate pool} of users who follow or are followed by known trolls. For inference, we ensembled the 10 models from 10-fold cross-validation by averaging predicted probabilities. The model initially flagged 25{,}381 users as suspected trolls. We excluded 1{,}383 users who had zero troll comments, leaving 23{,}998 users (0.59\% of 4,047,680) in the final set of model-detected accounts. We refer to these accounts as \emph{model-detected trolls}, but they should be interpreted as algorithmic predictions rather than independently verified state actors.

Among these users, 23{,}993 appeared in the article pool, 986 appeared in the follow-network pool, and 981 appeared in both pools. Only 5 model-detected trolls were found exclusively via the follow-network pool, whereas 23{,}012 were found exclusively in the article pool. This distribution suggests that, although proximity in the follow network to known trolls is highly informative, as reflected in the feature-importance results in Figure~\ref{fig:combined_ml_results}(a), the detector does not rely solely on the follow graph. It also incorporates signals from the explainable troll-content classifier and additional user-level behavioral features.

Table~\ref{tab:stance_examples} shows the translated examples for each target country category in comments posted by model-detected trolls. The \textit{Condemning Korea} example frames conscription as ``state-led slavery,'' while \textit{Condemning Rival} criticizes U.S. dollar dominance and portrays the United States as dependent on MNS. In contrast, \textit{Praising MNS} depicts the major neighboring state (i.e., troll origin) as a role model for Korea, emphasizing MNS's socialist system and economic rise. Finally, \textit{Praising Partner} derogates Ukrainian leadership as a ``comedy-idiot'' while glorifying Putin as a ``hero.'' Together, these examples illustrate the narrative styles produced by model-detected trolls. We further examine these patterns through systematic behavioral analysis in the following section.

\subsection{Characterizing Model-Detected Trolls}
To examine whether model-detected trolls exhibit troll-like behavioral patterns, we conduct comparative analyses of 23{,}998 model-detected trolls and 4 million non-troll users.

\paragraph{Time-series Evaluation.} Prior work~\cite{saeed2022trollmagnifier} validates troll detection by demonstrating that flagged accounts exhibit stronger temporal synchronization with known trolls than baseline users do.
Following this strategy, we evaluate temporal synchronization by comparing the activity time series of known trolls, model-detected trolls, and non-troll users. 
To capture temporal concentration, for each group we construct an activity share series, where each time interval is expressed as the percentage of the group's total comments over the full observation period.
We then compute Pearson correlations over a common observation window. Model-detected trolls exhibit consistently strong alignment with known trolls across all temporal resolutions, with daily $r=0.8144$, weekly $r=0.8354$, and monthly $r=0.8464$. In contrast, non-troll users show weaker alignment, with daily $r=0.6509$, weekly $r=0.6829$, and monthly $r=0.6996$.
Some baseline correlation for non-troll users is expected because our article-based candidate pool conditions on overlapping news exposure (i.e., commenting on the same articles as known trolls), which can increase temporal synchronization.
However, model-detected trolls show a consistently higher degree of synchronization.

To test whether the correlation between model-detected trolls and known trolls is significantly larger than that between non-troll users and known trolls, we perform a statistical comparison of dependent correlations that share the same reference series. We apply a Fisher $r$-to-$z$ transformation and use a Williams test for comparing correlations with an overlapping variable~\cite{williams1959comparison}.
The difference is statistically significant at all resolutions (daily $z=67.18$, weekly $z=28.54$, monthly $z=14.98$, all $p<0.001$), indicating that model-detected trolls remain significantly more synchronized with known trolls than non-troll users, beyond what is expected from shared news exposure.

\begin{table}[t]
\centering
\small
\caption{User-level average percentages computed from content-level predictions across six classes for model-detected trolls and non-troll users. Model-detected trolls consistently exhibit higher prediction rates across all six classes considered important within the hierarchical detection framework. ``MNS'' refers to \textit{a major neighboring state}.}

\label{tab:mean_class_prediction}
\begin{tabular}{ccc}
\toprule
\textbf{Class} & \textbf{Model-Detected Trolls} & \textbf{Non-Trolls} \\  \midrule
Troll Comments          & 67.19\% & 22.35\% \\
Foreign State-Suspected & 81.25\% & 31.24\% \\
Condemning Korea        & 57.16\% & 18.21\% \\
Condemning Rival       & 1.85\% & 0.53\% \\
Praising MNS  & 1.11\%  & 0.40\% \\
Praising Partner       & 0.004\% & 0.0004\% \\
\bottomrule
\end{tabular}
\end{table}

\paragraph{Content-Level Signals.} We used predictions from an explainable content-level model across six troll-related classes. Table~\ref{tab:mean_class_prediction} shows that model-detected trolls exhibit higher average prediction scores than non-troll users across all troll-related classes. We conducted a statistical analysis to examine whether user-level prediction scores differ between model-detected trolls and non-troll users. Levene's tests indicated violations of the equal variance assumption across all six classes (\textit{p} < 0.001), motivating the use of Welch's \textit{t}-tests, which are robust to heteroscedasticity. Prediction scores were significantly and consistently elevated for model-detected trolls across all classes (\textit{Troll Comments}: \textit{t} = 212.07; \textit{Foreign State-Suspected}: \textit{t} = 220.69; \textit{Condemning Korea}: \textit{t} = 185.14; \textit{Condemning Rival}: \textit{t} = 42.84; \textit{Praising MNS}: \textit{t} = 25.88; \textit{Praising Partner}: \textit{t} = 3.41; all \textit{p} < 0.001) and remained significant after Bonferroni correction ($\alpha = 0.0083$), indicating robustness across multiple tests. These results indicate that the user-level model effectively captures systematic variation between model-detected trolls and non-troll accounts within the predicted troll-related classes (see details on user-level traits in Appendix~\ref{sec:comparative_troll}).

\begin{figure}[t]
\centering
\includegraphics[width=0.48\textwidth]{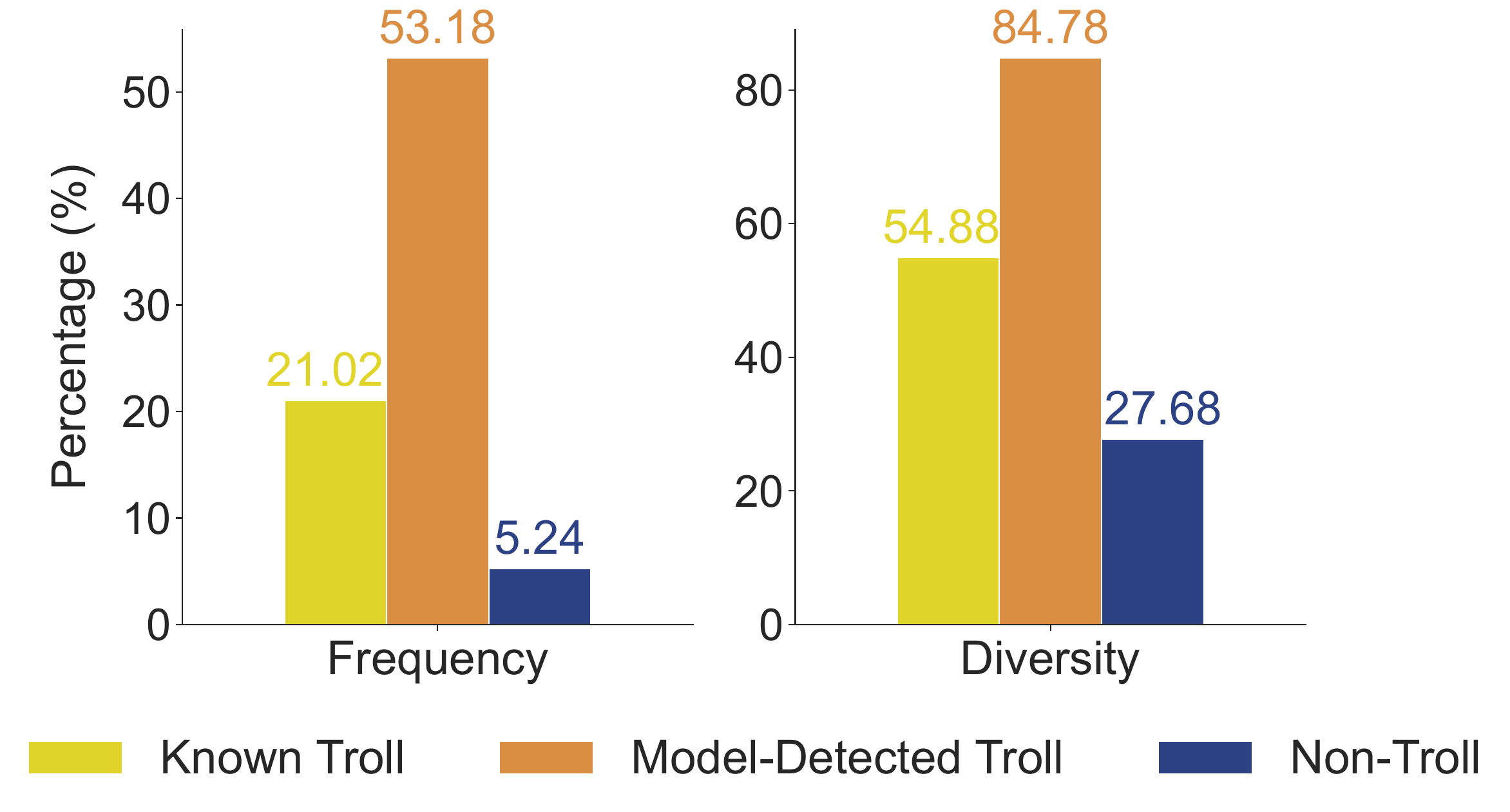}
\caption{MNS language frequency and diversity by user group, computed from tokens within spans predicted as \textit{Foreign State-Suspected} (excluding the Writing Region token). Frequency: \% of MNS-language tokens among all tokens in these spans. Diversity: \% of unique MNS-language tokens among MNS-language tokens in these spans.}
\label{fig:mns_language}
\end{figure}

\paragraph{Span-Level Linguistic Evidence.} We examined the prevalence of MNS language expressions at the span level. For comparison, we removed the Writing Region token, which appeared frequently in all groups. All measures were computed from tokens extracted within spans predicted as \textit{Foreign State-Suspected}. As shown in Figure~\ref{fig:mns_language}, model-detected trolls heavily use culture-specific markers and jargon associated with the major neighboring state, whereas non-troll users rarely exhibit this pattern. Chi-squared tests confirmed significant differences in MNS language frequency and diversity across groups ($p < 0.001$). Post-hoc one-tailed Z-tests for proportions further showed that both known and model-detected troll groups exhibit significantly higher frequency (Known vs. Non-Troll: $Z=31.20$; Model-Detected vs. Non-Troll: $Z=136.67$) and diversity (Known vs. Non-Troll: $Z=13.55$; Model-Detected vs. Non-Troll: $Z=62.42$), all with $p < 0.001$ compared to non-troll users. These results indicate that model-detected trolls tend to exhibit higher frequency and diversity of MNS language usage, suggesting clear linguistic differences from non-troll users.

These time-series, content-level, and span-level analyses indicate that model-detected trolls differ systematically from the broader user baseline. Accounts flagged by the model exhibit temporal activity patterns that align more closely with known trolls than with typical non-troll users, have higher average user-level prediction scores across the six troll-related classes, and display distinctive linguistic indicators at the span level. Further comparative observations regarding these model-detected trolls and baseline users are provided in Appendix~\ref{sec:comparative_troll}.

\begin{figure*}[t]
\centering
\includegraphics[width=0.92\textwidth]{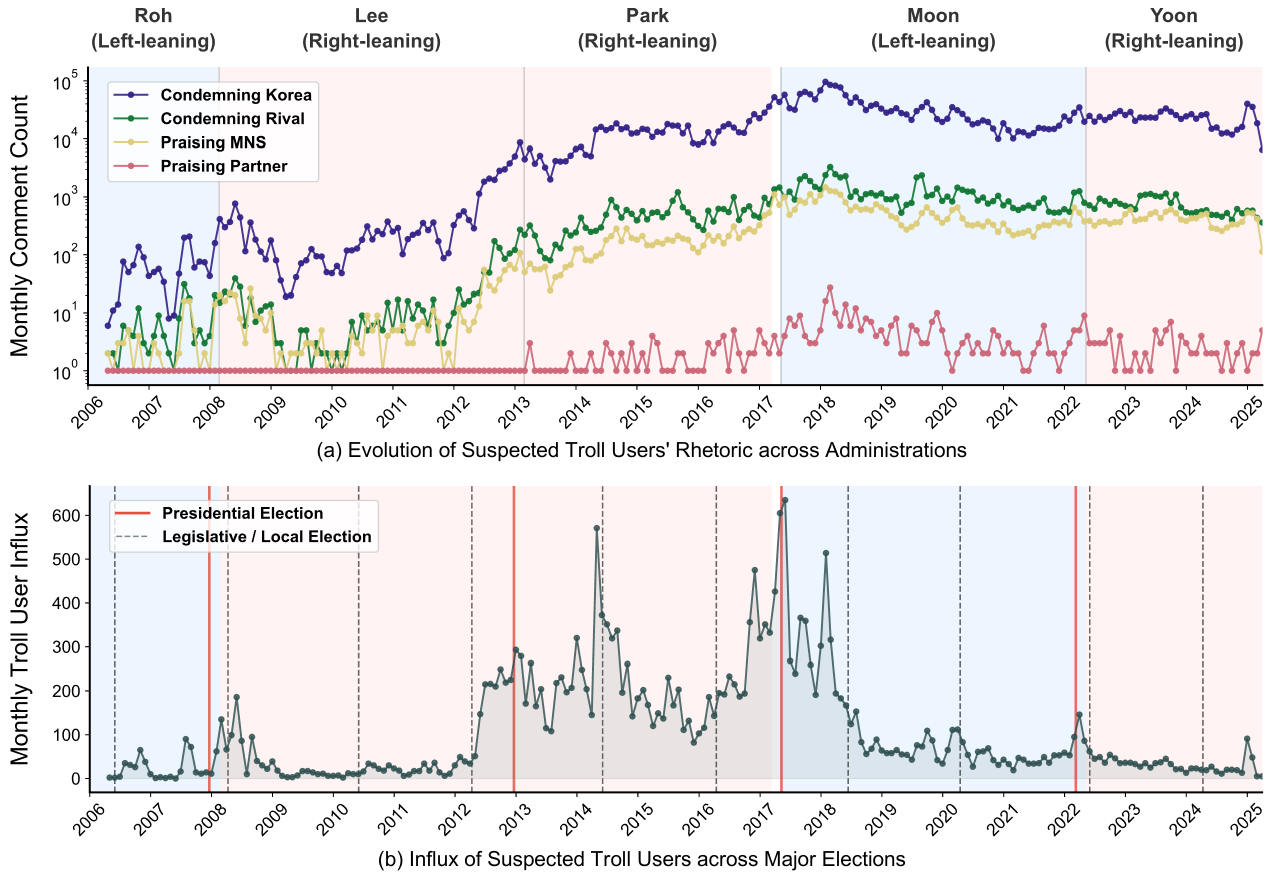}
\caption{Temporal patterns of troll activity from 2006 to 2025. \textbf{(a)} Evolution of comment volume across different rhetorical categories. Background colors indicate governing administrations (blue = left-leaning; red = right-leaning), with presidents' last names shown at the top. The unshaded period in 2017 represents the transitional period following impeachment before the new administration. \textbf{(b)} Monthly influx of newly identified troll-like users, calculated based on each user's first comment date. Red vertical lines mark presidential elections, while black dashed lines indicate legislative and local elections.}
\label{fig:temporal_trend}
\end{figure*}



\section{Strategy Analysis}
\label{sec:strategy_analysis}

\begin{figure*}[t]
    \centering
    \includegraphics[width=\textwidth]{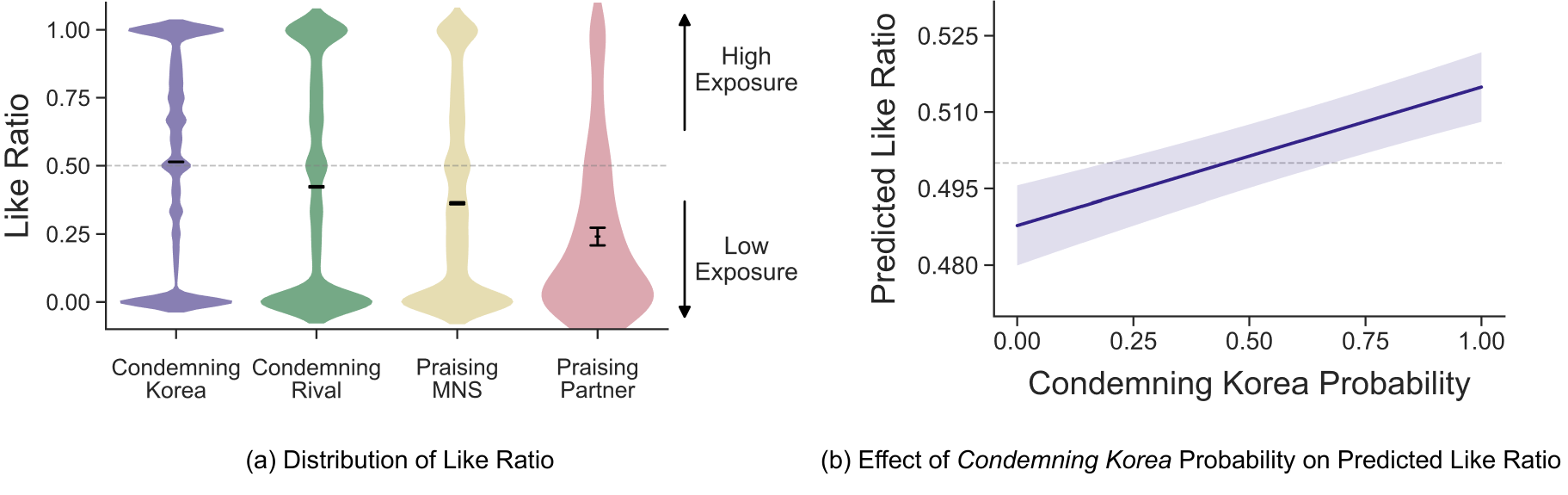} 
    \caption{Like ratio by rhetorical strategy. \textbf{(a)} Distribution of like ratios across strategies with the neutral threshold (like = dislike) shown as a dashed line; \textit{Condemning Korea} is the only strategy whose mean exceeds this threshold.  Error bars denote 95\% confidence intervals around the mean. \textbf{(b}) Predicted like ratio from a fractional logit model as a function of the \textit{Condemning Korea} probability, with user-clustered standard errors and 95\% confidence intervals. }
\label{fig:combined_analysis}
\end{figure*}

For strategy analysis, we combine the verified seed set of 70 known trolls with the 23{,}998 model-detected accounts, yielding a consolidated pool of $24{,}068$ {suspected troll accounts}. To focus on influence-related behavior, we restrict the analysis to comments classified as troll comments (4{,}123{,}962), rather than all comments authored by these users (15{,}054{,}072). 
This restriction helps isolate campaign-relevant messaging from broader, non-strategic discussion (e.g., memes). We characterize campaign-level strategies along three dimensions: (i) longitudinal trend analysis, (ii) comment visibility by rhetoric, and (iii) targeted entities in the most visible comments.

\subsection{Longitudinal Trend Analysis} 
\paragraph{20-Year Rhetorical Asymmetry.} As shown in Figure~\ref{fig:temporal_trend} (a), we analyze the longitudinal trajectory of comments authored by suspected troll accounts over a 20-year period (monthly counts; log scale). A primary observation is the marked asymmetry in rhetorical emphasis: regardless of administration changes, suspected troll accounts posted substantially more \textit{condemning} content than \textit{praising} content. ``Condemning Korea'' remains the dominant narrative throughout the entire period. Given that Naver News is a major domestic news platform, this pattern suggests efforts to influence domestic discourse through adversarial framing rather than overt promotion of external narratives.

All rhetorical categories show upward trends from 2017, reaching their maximum volumes around 2018. In 2018, ``Condemning Korea'' peaks at 675,107 comments, followed by ``Condemning Rival'' (20,254), ``Praising MNS (major neighboring state)'' (10,345), and ``Praising Partner'' (108). This surge temporally coincides with two major developments: (1) the 2017 institutionalization of military-civil fusion and cognitive-domain operations by MNS, which has been discussed as enabling more coordinated psychological and informational campaigns~\cite{sullivan2025howchinafights}, and (2) the escalating diplomatic conflict between Korea and MNS over the deployment of the US THAAD missile defense system, which intensified to the point of economic retaliation.

\paragraph{Election-Oriented Entry Patterns.} Motivated by prior evidence that coordinated influence activity intensifies around elections~\cite{ssci2019report, howard2018ira}, we examine whether and to what extent suspected trolls join the platform around elections. Figure~\ref{fig:temporal_trend} (b) plots the influx of newly detected troll users, computed based on their first observed comment date, with major presidential, legislative, and local elections annotated. New account activity increases around these pivotal political cycles. The largest spike occurs around the May 2017 presidential election following President Park's impeachment, with 635 newly appearing suspected troll accounts. 

To evaluate the statistical significance of this pattern, we analyze user influx at a weekly resolution, balancing the aggregation bias of monthly counts against the sparsity of daily data. Election periods are defined as weeks within $\pm 30$ days of any major election and are compared against non-election weeks using the Mann-Whitney U test. The average weekly influx of flagged troll-like users is 34.19 during election periods compared to 22.61 during non-election periods (+51.23\%; $p<0.01$). A large cohort of non-troll users (4M) also shows a higher influx during election periods (4{,}828.11 vs.\ 3{,}893.42; +24.01\%; $p<0.05$), suggesting that elections generally attract new platform participation. 

To assess whether suspected trolls concentrate disproportionately around elections beyond this general influx, we compute \textit{penetration density}, defined as the weekly fraction of newly appearing users classified as trolls. Penetration density increases from 0.82\% in non-election periods to 0.91\% during election periods (+10.98\%; $p<0.05$, Mann-Whitney U). These results indicate that troll-like accounts show elevated entry rates around elections, consistent with election-oriented mobilization patterns documented in prior work~\cite{ssci2019report, howard2018ira}.

\subsection{Comment Visibility by Rhetoric Strategy} We analyzed which rhetorical strategies employed by suspected troll accounts were most likely to achieve visibility among Naver News users. On this platform, comment visibility is determined by the like ratio, calculated as 
\begin{equation}
    \frac{\text{likes}}{(\text{likes} + \text{dislikes})}
\end{equation} 
with higher ratios elevating comments to the top of the thread. Comments with zero engagement (both likes and dislikes equal to zero) were assigned a like ratio of 0, as they remain invisible at the bottom of threads, similar to comments receiving only dislikes.

\paragraph{Cross-Strategy Statistical Comparison.}  Figure~\ref{fig:combined_analysis} (a) shows the like ratios associated with different rhetorical strategies. Among troll strategies, condemning rhetoric achieved higher like ratios on average than praising rhetoric. \textit{Condemning Korea} was the only strategy to exceed the neutral threshold of 0.5 (mean $= 0.514$), indicating it received more likes than dislikes on average and was most likely to achieve top-thread visibility. In contrast, \textit{Condemning Rival} (mean $= 0.422$), \textit{Praising MNS} (mean $= 0.362$), and \textit{Praising Partner} (mean $= 0.241$) all fell below the 0.5 mark. Because like-ratio distributions were non-normal across all categories (Shapiro-Wilk tests, $p < 0.001$), we applied a Kruskal-Wallis test, which revealed significant differences among categories ($H = 13101.53, p < 0.001$). Post-hoc Dunn tests with Bonferroni correction confirmed that all pairwise contrasts were significant ($p < 0.001$). 

These results indicate that the visibility of troll content varied significantly by the emotional target of the message, with condemning rhetoric consistently outperforming praising rhetoric in terms of user engagement. This pattern holds across political administrations (see  Figure~\ref{fig:like_ratio_across_administration} in the Appendix).  \textit{Condemning Korea} ranked first in four out of the five periods, with the sole exception occurring under the Lee administration, where \textit{Condemning Rival} showed the highest like ratio.  Across all regimes, praising rhetoric persistently received lower engagement than condemning rhetoric.

\paragraph{Impact of Rhetorical Intensity on Visibility.} To further examine whether the intensity of each rhetorical strategy influenced visibility, we conducted a regression analysis. Following prior work~\cite{kim2024moral}, we operationalized rhetorical intensity as the softmax probability from our ELECTRA-based content-level classifier's output logits. This probability represents the model's confidence in classifying a comment into a specific rhetorical category, which we interpret as the strength of that rhetorical strategy in the content. We performed a regression with the four rhetorical intensity scores (\textit{Condemning Korea}, \textit{Condemning Rivals}, \textit{Praising MNS}, \textit{Praising Partners}) as independent variables and the like ratio as the dependent variable. Because the dependent variable ranges between 0 and 1 with substantial mass at both boundaries (30.44\% at 0 and 21.32\% at 1), we employed a Fractional Logit Model~\cite{papke1996econometric}. Year and month fixed effects were included to account for temporal and seasonal variability.

Table~\ref{tab:logit_results} shows the effects of rhetorical intensity on comment like ratios.  Among the strategies, only \textit{Condemning Korea} significantly boosted engagement ($\beta = 0.1086$, $P < 0.001$), directly enhancing content visibility. This relationship is illustrated in  Figure~\ref{fig:combined_analysis}(b): as the intensity of \textit{Condemning Korea} increases, it crosses the neutral threshold (where likes equal dislikes) at a probability of 0.45.  In contrast, both \textit{Condemning Rival} ($\beta = -0.5093, P < 0.001$) and \textit{Praising MNS} ($\beta = -1.0888, P < 0.001$) significantly suppressed like ratios. While \textit{Praising  Partner} also yielded a negative coefficient ($\beta = -0.2072$), its effect was not statistically significant. The negative coefficient for \textit{Praising MNS} suggests that this narrative strategy triggers intense audience antipathy, rendering such content the least likely to achieve organic visibility.
\begin{table}[t]
\caption{Fractional Logit estimates: Impact of rhetoric intensity on Like Ratio ($N$=4,123,692). All models control for year and month fixed effects. Significance levels: $^{***}p<.001$ (\colorbox{green!25}{positive}, \colorbox{red!25}{negative}).}
\vspace{-2mm}
\centering
\small
\frenchspacing
\renewcommand{\arraystretch}{1.3}
\begin{tabular}{lc}
\toprule
\textbf{Independent Variable (Intensity)} & \textbf{Like Ratio (Coef.)} \\
\midrule
Condemning Korea & \cellcolor{green!25}0.1086$^{***}$ \\
Condemning Rival & \cellcolor{red!25}-0.5093$^{***}$ \\
Praising MNS & \cellcolor{red!25}-1.0888$^{***}$ \\
Praising Partner & \cellcolor{red!25}-0.2072 \\
\bottomrule
\end{tabular}
\label{tab:logit_results}
\end{table}

\subsection{Targeted Entities in Condemning Korea} 
We conducted a deeper analysis of the \textit{Condemning Korea} strategy, which showed the highest engagement potential over the 20-year period and the greatest likelihood of achieving top-thread visibility. Our explainable content-level troll detection framework provides span-level rationales that explicitly mark who the \textit{Condemning Korea} rhetoric targets, enabling us to identify the frequently targeted entities.

\begin{figure}[!t]
\centering
\hspace{-1mm}
\includegraphics[width=0.5\textwidth]{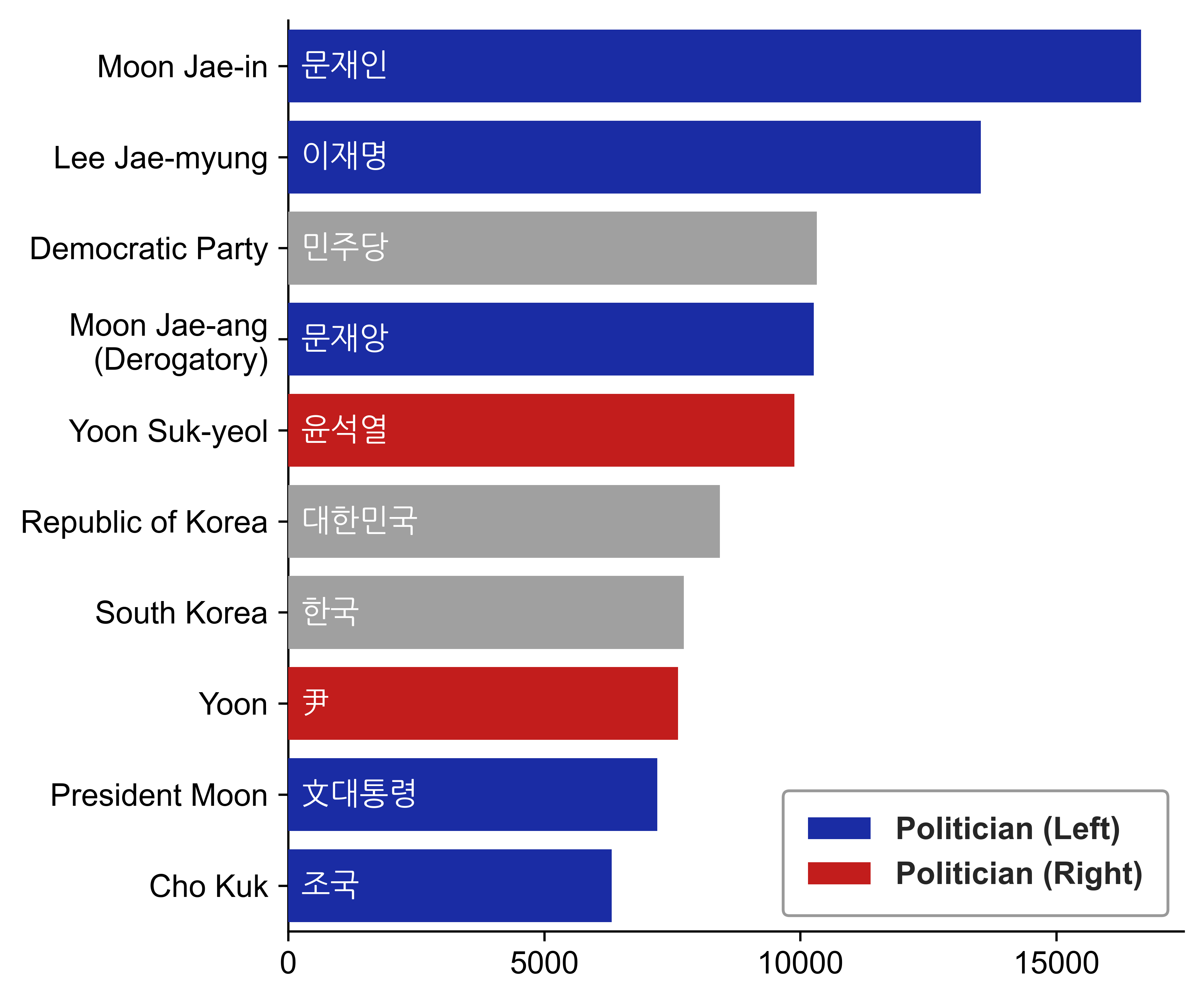}
\caption{Top-10 targets in \textit{Condemning Korea} comments with the highest like ratio (=1). Bars show the frequency of target spans extracted by the content‑level detector. Colors denote target type: political figures by ideological camp (left‑leaning vs. right‑leaning), and non‑figure targets in gray. Text inside each bar shows original Korean spans with English translations on the left.}
\label{fig:span_frequency}
\end{figure}

Figure~\ref{fig:span_frequency} presents the ten most frequent target spans in \textit{Condemning Korea} comments with the highest like ratio (=1). Seven of the top ten targets are political leaders. The most frequently mentioned figures are \textit{Moon Jae-in} (liberal-leaning former president; 16,651 mentions), \textit{Lee Jae-myeong} (liberal-leaning presidential candidate; 13,522 mentions), \textit{Yoon Suk-yeol} (conservative president; 9,887 mentions), and \textit{Cho Kuk} (liberal-leaning politician; 6,314 mentions). The targets span both liberal and conservative camps, consistent with prior work suggesting that troll activity tends to amplify polarization~\cite{arif2018acting}. Targeting high-salience political figures may be associated with higher visibility and engagement in comment threads. Appendix Table~\ref{tab:target_span_full_list_updated} shows that presidents, major candidates, and other elite-associated figures frequently appear among the top-10 target spans in high-visibility \textit{Condemning Korea} comments across administrations. The incumbent president consistently appears among the top-10 targets regardless of ideological orientation. 
This trend is more pronounced in the Moon and Yoon administrations, where political figures account for a larger share of the most frequent target spans.

These targets range from political leaders (acting as representatives of parties and the state) to political parties and the nation itself, aligning with core attributes of moralized content as discourse oriented toward units larger than the individual (e.g., society or culture)~\cite{haidt2003moral}. The concentration of these collective-level targets among the most visible comments is notable, consistent with prior findings that moralized messaging captures attention and can spread through online networks~\cite{brady2017emotion, park2026moral}. By achieving high visibility, these comments disproportionately expose audiences to rhetoric centered on systemic or collective concerns rather than private matters, which prior research links to more polarized discourse~\cite{brady2017emotion}.


\section{Discussion}
\label{sec:discussion}


\subsection{Findings and Stakeholder Reporting}
Over the course of nearly two decades, our analysis reveals that \textit{Condemning Korea} is both the dominant and most amplified narrative in influence campaigns from potentially the major neighboring state. This pattern aligns with cognitive‑warfare research suggesting that negative, antagonistic rhetoric is particularly effective at capturing user attention~\cite{claverie2022cognitive}. 
To support immediate verification and mitigation, the list of 24,068 suspected troll accounts has been shared with Naver News and the Institute for National Security Strategy in South Korea, the organization that released the initial troll label dataset used as our seeds. These shared data provide platform stakeholders and policy observers with the actionable insights required to address systemic account irregularities and uphold the integrity of public discourse.

\subsection{Implications for Online Platforms}
The proposed explainable content-level detector provides a concrete mechanism for strengthening platform governance. Moderation decisions such as warnings, down-ranking, or account restrictions are frequently contested because affected users receive little insight into why an action was taken~\cite{myers2018censored, jhaver2019does}. By producing hierarchical label predictions, our model can surface specific textual evidence underlying moderation outcomes, thereby supporting moderation and review workflows and making enforcement decisions more transparent, interpretable, and contestable.

Another consideration is deployment at scale. Although LLMs are used during training through knowledge distillation, our inference relies on a smaller model ($\approx 0.1\text{B}$) while maintaining strong predictive performance. This lightweight design is suitable for production environments, where latency and cost constraints typically limit the use of large models.

Our findings suggest that adversarial adaptation may not be cost-free, even if the model's decision logic becomes known. In this setting, the cues that drive comment visibility are closely aligned with the rhetorical signals captured by our theory-grounded, explainable detector. To evade detection, attackers would need to weaken these rhetorical cues; however, because our regression analysis shows that these cues are associated with visibility (Table~\ref{tab:logit_results}), evasion may come at the expense of reduced reach and influence~\cite{brady2017emotion,kim2024moral}. In other words, evasion attempts can create a trade-off between bypassing detection and maintaining strategic effectiveness.

\subsection{Implications for Defensive Prioritization}
Effective responses to information operations require more than detection; they demand strategic decisions about intervention and resource prioritization~\cite{wardle2017information, starbird2019disinformation}. As harmful narratives may become harder to correct once they gain widespread visibility~\cite{starbird2019disinformation,giglietto2019fake}, limited defensive resources should be allocated to timely and targeted interventions.

Our findings suggest that prioritization should account for both when to monitor and what to monitor. The concentration of suspected troll entry around major elections indicates that election periods are especially important windows for heightened scrutiny. At the same time, condemning moral-emotional rhetoric has grown in both prevalence and visibility over two decades, and condemnation of political figures across the political spectrum was particularly likely to gain exposure. These patterns suggest that not all suspicious content carries the same amplification risk, as engagement-based ranking dynamics can preferentially amplify moral-emotional expression~\cite{brady2017emotion,park2026moral}. Platforms and observatories can allocate additional review capacity during major election windows and prioritize the verification and, where appropriate, debunking of suspicious messages that employ condemning rhetoric and target political figures, enabling intervention before such messages achieve widespread reach.



\section{Conclusion}
\label{sec:conclusion}
This study presents a scalable framework for detecting adversarial accounts within online news comment sections. Our approach offers \textit{explainable} rationales via a hierarchical content-level classifier that evaluates foreign state–suspected origin, moral emotions, and target entities of influence operations. By aggregating these outputs into user-level behavioral features from longitudinal data, we track how underlying narrative strategies evolved and gained visibility on a South Korean platform over nearly twenty years.

Our findings reveal a persistent pattern: rhetoric focused on moral condemnation dominates troll activity and consistently achieves higher visibility than content praising foreign agendas. The number of new troll-like accounts increases around elections, indicating a strategic responsiveness to major political events. Moreover, span-level evidence shows that domestic political leaders from both liberal and conservative camps are repeatedly targeted. This cross-partisan targeting demonstrates that these operations prioritize intensifying polarization and societal distrust rather than advancing a specific partisan position.

Our findings offer a unique empirical view into the real-world evolution of influence operations within South Korean news comments. They hold significant implications for platform governance. Because individual news readers lack this global view, they cannot easily discern which isolated comments are part of a coordinated foreign influence campaign. The use of span-level rationales provides interpretable evidence for content moderation, while the strategy analysis indicates which narratives are most likely to surface and persist. Together, these insights can inform platform policy and help prioritize defensive attention toward the most engagement-effective tactics.
\clearpage





\section*{Ethical Considerations}
\paragraph{Data Collection and Privacy.}
All data were collected from the public-facing Naver News platform in accordance with the platform's terms of service, consisting of public comments and associated metadata. To minimize the risk of user re-identification, usernames are partially masked in all public-facing outputs, including the text examples in this paper and the open repository. Furthermore, computational analyses were conducted entirely on secure institutional infrastructure. We also strictly focus on aggregate behavioral patterns rather than individual profiles. Accordingly, no attempts were made to infer sensitive personal attributes, such as political affiliation, demographic characteristics, or real-world identities.

\paragraph{Data Removal.}
Although the public release is partially masked, we retain the full usernames on access-restricted internal infrastructure, which lets us process removal requests. The Zenodo release includes a bilingual Korean-English guide explaining how individuals can locate their comments using the visible username prefix and comment text to submit a removal request, if needed. For account removal, individuals can provide evidence of account ownership, such as a screenshot of their Naver profile displaying the full username. For comment removal, individuals can specify the relevant comment URL. Upon verification, the corresponding rows will be removed from our maintained research dataset and all future public releases.

\paragraph{Stakeholders and Potential Harms.}
This research has operational implications for suspected troll accounts, general news commenters, civic organizations, and the platform itself. For model-detected accounts, the primary risk involves misclassification and subsequent reputational harm if they are incorrectly labeled as state actors. To mitigate this risk, we consistently use the terms ``suspected'' and ``model-detected,'' report aggregate behavioral patterns rather than individual accusations, and withhold unmasked account lists from the public release. We mask all general commenter usernames across public outputs to prevent unwarranted associations. For civil-society groups, we report prevalence and trends rather than actor attribution or operational recommendations against specific countries. For the platform, we used only its public API within the terms of service and rate limits and shared our findings for follow-up verification.

\paragraph{Researcher Well-being.}
The research required repeated exposure to hostile, politically charged comments, which can impose a psychological burden on researchers. All annotation and close reading were conducted by research team members rather than external or paid annotators. To mitigate potential harm, we used pre-annotation risk briefings with opt-out provisions, regular check-ins, team-based review of the most difficult content rather than solo review, access to institutional counseling, and pause requests without exception.

\paragraph{Research Justification and Public Interest.}
We conducted this study after carefully evaluating potential risks and determining that the public utility of tracking long-term influence operations on Naver significantly outweighed those concerns. We framed this work as defensive measurement research designed to support digital governance, independent scrutiny, and civil-society awareness. To mitigate potential misuse, we avoid attributing model-detected accounts to real-world identities or independently verified state actors, releasing only aggregate analyses and partially masked data. These safeguards allow the findings to inform defensive strategies while minimizing the risks of harassment, retaliation, or the direct targeting of individual users.

\section*{Open Science}
To support reproducibility and future research, our complete dataset is publicly available on Zenodo at \url{https://doi.org/10.5281/zenodo.20257085}. This repository includes (1) a near 20-year corpus of 112 million news comments with partially masked user identifiers to protect privacy and (2) a bilingual Korean-English administrative guide detailing the user data removal and opt-out verification process.


\section*{Acknowledgments}
We thank Carmela Troncoso and researchers at the MPI-SP for their valuable feedback. This work was supported by the Hyundai Motor Chung Mong-Koo Foundation, the IITP grant (RS-2024-00441762), and the NRF grant (RS-2022-00165347) funded by the Korean government (MSIT).

\bibliographystyle{plain}
\bibliography{references}

@article{kelton2019australia,
  title={Australia, the utility of force and the society-centric battlespace},
  author={Kelton, Maryanne and Sullivan, Michael and Bienvenue, Emily and Rogers, Zac},
  journal={International Affairs},
  year={2019},
  publisher={Oxford University Press}
}

@misc{lee2021kcelectra,
  author = {Junbum Lee},
  title = {{KcELECTRA}: Korean Comments {ELECTRA}},
  year = {2021},
  publisher = {GitHub},
  journal = {GitHub repository},
  howpublished = {\url{https://github.com/Beomi/KcELECTRA}}
}

@inproceedings{park2021klue,
  title        = {{KLUE}: Korean Language Understanding Evaluation},
  author       = {Park, Sungjoon and Moon, Jihyung and Kim, Sungdong and others},
  booktitle    = {35th Conference on Neural Information Processing Systems Track on Datasets and Benchmark},
  year         = {2021}

}

@inproceedings{loshchilov2019adam,
  author       = {Ilya Loshchilov and
                  Frank Hutter},
  title     = {Decoupled weight decay regularization},
  booktitle = {International Conference on Learning Representations},
  year         = {2019}
}

@inproceedings{lee2020kcbert,
  title={{KcBERT}: Korean Comments {BERT}},
  author={Lee, Junbum},
  booktitle={Proceedings of the 32nd Annual Conference on Human and Cognitive Language Technology},
  year={2020}
}

@inproceedings{kim2024moral,
  title={How do moral emotions shape political participation? A cross-cultural analysis of online petitions using language models},
  author={Kim, Jaehong and Jeong, Chaeyoon and Park, Seongchan and Cha, Meeyoung and Lee, Wonjae},
  booktitle={Findings of the Association for Computational Linguistics},
  year={2024}
}

@techreport{bernal2020cognitive,
  title        = {Cognitive Warfare: An Attack on Truth and Thought},
  author       = {Bernal, A. and Carter, C. and Singh, I. and Cao, K. and Madreperla, O.},
  year         = {2020},
  institution  = {NATO Innovation Hub and Johns Hopkins University Applied Physics Laboratory},
  url          = {https://innovationhub-act.org/wp-content/uploads/2023/12/Cognitive-Warfare.pdf},
}

@misc{labelstudio,
  title={{Label Studio}: Data labeling software},
  howpublished = {\url{https://github.com/HumanSignal/label-studio}},
  author={Maxim Tkachenko and Mikhail Malyuk and Andrey Holmanyuk and Nikolai Liubimov},
  year={2020--2025},
}

@inproceedings{demszky2020goemotions,
  title = "{G}o{E}motions: A Dataset of Fine-Grained Emotions",
    author = "Demszky, Dorottya  and
      Movshovitz-Attias, Dana  and
      Ko, Jeongwoo  and
      Cowen, Alan  and
      Nemade, Gaurav  and
      Ravi, Sujith",
  booktitle = "Proceedings of the 58th Annual Meeting of the Association for Computational Linguistics",
  year={2020}
}

@article{arif2018acting,
  title={Acting the part: Examining information operations within\# {BlackLivesMatter} discourse},
  author={Arif, Ahmer and Stewart, Leo Graiden and Starbird, Kate},
  journal={Proceedings of the ACM on Human-Computer Interaction},
  year={2018},
  publisher={ACM New York, NY, USA}
}

@misc{achiam2023gpt,
  title={{GPT-4} Technical Report},
  author={Josh Achiam and Steven Adler and Sandhini Agarwal and others},
  year={2023},
  eprint={2303.08774},
  archivePrefix={arXiv},
  url={https://arxiv.org/abs/2303.08774}
}

@incollection{claverie2022cognitive,
  title     = {``Cognitive warfare'': The advent of the concept of ``cognitics'' in the field of warfare},
  author={Claverie, Bernard and Du Cluzel, Fran{\c{c}}ois},
  booktitle = {Cognitive Warfare: The Future of Cognitive Dominance},
  year={2022},
  publisher={NATO Collaboration Support Office}
}

@techreport{sullivan2025howchinafights,
  author       = {Ian Sullivan},
  title        = {How {China} Fights in Large-Scale Combat Operations},
  institution  = {U.S. Army Training and Doctrine Command (TRADOC G-2)},
  year         = {2025},
  url          = {https://www.army.mil/article/285395/how_china_fights_in_large_scale_combat_operations}
}

@article{ding2025span,
  title={Span-Oriented Information Extraction: A Unified Framework},
  author={Ding, Yifan and Yankoski, Michael and Weninger, Tim},
  journal={ACM SIGKDD Explorations Newsletter},
  year={2025},
  publisher={ACM New York, NY, USA}
}

@inproceedings{hasanain-etal-2024-large,
    title = "Large Language Models for Propaganda Span Annotation",
    author = "Hasanain, Maram  and
      Ahmad, Fatema  and
      Alam, Firoj",
    booktitle = "Findings of the Association for Computational Linguistics: EMNLP",
    year = "2024",
}

@inproceedings{brown2020language,
  title={Language models are few-shot learners},
  author={Brown, Tom and Mann, Benjamin and Ryder, Nick and others},
  journal={Advances in neural information processing systems},
  booktitle={Proceedings of the 34th International Conference on Neural Information Processing Systems},
  year={2020}
}

@article{field2022analysis,
  title={An analysis of emotions and the prominence of positivity in\# {BlackLivesMatter} tweets},
  author={Field, Anjalie and Park, Chan Young and Theophilo, Antonio and Watson-Daniels, Jamelle and Tsvetkov, Yulia},
  journal={Proceedings of the National Academy of Sciences},
  year={2022},
  publisher={National Academy of Sciences}
}

@book{csis2023tenfortaiwan,
  title={Building International Support for Taiwan},
  author={Blanchette, Jude and Hass, Ryan and McElwee, Lily},
  year={2024},
  publisher={JSTOR}
}

@incollection{haidt2003moral,
  title     = {The moral emotions},
  author    = {Haidt, Jonathan},
  booktitle = {Handbook of Affective Sciences},
  publisher = {Oxford University Press},
  year      = {2003}
}

@article{van2024social,
  title={Social media and morality},
  author={Van Bavel, Jay J and Robertson, Claire E and Del Rosario, Kareena and Rasmussen, Jesper and Rathje, Steve},
  journal={Annual Review of Psychology},
  year={2024},
  publisher={Annual Reviews}
}

@article{brady2020mad,
  title={The {MAD} model of moral contagion: The role of motivation, attention, and design in the spread of moralized content online},
  author={Brady, William J and Crockett, Molly J and Van Bavel, Jay J},
  journal={Perspectives on Psychological Science},
  year={2020},
  publisher={Sage Publications Sage CA: Los Angeles, CA}
}

@article{brady2017emotion,
  title={Emotion shapes the diffusion of moralized content in social networks},
  author={Brady, William J and Wills, Julian A and Jost, John T and Tucker, Joshua A and Van Bavel, Jay J},
  journal={Proceedings of the National Academy of Sciences},
  year={2017},
  publisher={National Academy of Sciences}
}

@inproceedings{zampieri-etal-2019-predicting,
    title = "Predicting the Type and Target of Offensive Posts in Social Media",
    author = "Zampieri, Marcos  and
      Malmasi, Shervin  and
      Nakov, Preslav  and
      Rosenthal, Sara  and
      Farra, Noura  and
      Kumar, Ritesh",
    booktitle = "Proceedings of the 2019 Conference of the North {A}merican Chapter of the Association for Computational Linguistics: Human Language Technologies",
    year = "2019",
}

@inproceedings{jeong-etal-2022-kold,
    title = "{KOLD}: {K}orean Offensive Language Dataset",
    author = "Jeong, Younghoon  and
      Oh, Juhyun  and
      Lee, Jongwon  and
      Ahn, Jaimeen  and
      Moon, Jihyung  and
      Park, Sungjoon  and
      Oh, Alice",
    booktitle = "Conference on Empirical Methods in Natural Language Processing",
    year = "2022",
}

@inproceedings{park2026moral,
title={Moral Outrage Shapes Commitments Beyond Attention: Multimodal Moral Emotions on {YouTube} in {Korea} and the {US}},
author={Park, Seongchan and Kim, Jaehong and Kim, Hyeonseung and Bin, Heejin and Moon, Sue and Lee, Wonjae},
booktitle={Proceedings of the {ACM} Web Conference},
year={2026}
}

@article{hung2022china,
  title={How {China}'s cognitive warfare works: a frontline perspective of {Taiwan}'s anti-disinformation wars},
  author={Hung, Tzu-Chieh and Hung, Tzu-Wei},
  journal={Journal of Global Security Studies},
  year={2022},
  publisher={Oxford University Press}
}

@article{simchon2022troll,
  title={Troll and divide: the language of online polarization},
  author={Simchon, Almog and Brady, William J and Van Bavel, Jay J},
  journal={PNAS Nexus},
  year={2022},
  publisher={Oxford University Press}
}

@article{han2025commenters,
  title={Commenters and lurkers: Navigating the two-step flow of communication in online news discourse},
  author={Han, Jiyoung},
  journal={New Media \& Society},
  year={2025},
  publisher={SAGE Publications Sage UK: London, England}
}

@article{schroeder2026malicious,
  title={How malicious {AI} swarms can threaten democracy},
  author={Schroeder, Daniel Thilo and Cha, Meeyoung and Baronchelli, Andrea and Bostrom, Nick and Christakis, Nicholas A and Garcia, David and Goldenberg, Amit and Kyrychenko, Yara and Leyton-Brown, Kevin and Lutz, Nina and others},
  journal={Science},
  year={2026},
  publisher={American Association for the Advancement of Science}
}

@misc{INSS2024ForeignInfluence,
  title        = {Current State of Foreign Influence Operations: Examples of Internet and Media Misuse},
  author  = {{Institute for National Security Strategy (INSS)}},
  year         = {2024},
  howpublished          = {\url{https://www.inss.re.kr/en/News/bbs/news_en_view.do?nttId=41037284}}
}

@article{giglietto2019fake,
  title={‘Fake news’ is the invention of a liar: How false information circulates within the hybrid news system},
  author={Giglietto, Fabio and Iannelli, Laura and Valeriani, Augusto and Rossi, Luca},
  journal={Current Sociology},
  year={2019},
  publisher={SAGE Publications Sage UK: London, England}
}

@article{starbird2019disinformation,
  title={Disinformation's spread: bots, trolls and all of us},
  author={Starbird, Kate},
  journal={Nature},
  year={2019},
  publisher={Nature Publishing Group}
}

@book{wardle2017information,
  title={Information disorder: Toward an interdisciplinary framework for research and policymaking},
  author={Wardle, Claire and Derakhshan, Hossein},
  year={2017},
  publisher={Council of Europe}
}

@inproceedings{recabarren2023strategies,
  title={Strategies and vulnerabilities of participants in {Venezuelan} influence operations},
  author={Recabarren, Ruben and Carbunar, Bogdan and Hernandez, Nestor and Shafin, Ashfaq Ali},
  booktitle={32nd USENIX Security Symposium},
  year={2023}
}

@inproceedings{lundberg2017unified,
  title={A unified approach to interpreting model predictions},
  author = {Lundberg, Scott M. and Lee, Su-In},
  booktitle = {Proceedings of the 31st International Conference on Neural Information Processing Systems},
  journal={Advances in neural information processing systems},
  year={2017}
}

@article{williams1959comparison,
  title={The comparison of regression variables},
  author={Williams, Evan J},
  journal={Journal of the Royal Statistical Society: Series B (Methodological)},
  year={1959},
  publisher={Wiley Online Library}
}

@article{myers2018censored,
  title={Censored, suspended, shadowbanned: User interpretations of content moderation on social media platforms},
  author={Myers West, Sarah},
  journal={New Media \& Society},
  year={2018},
  publisher={SAGE Publications Sage UK: London, England}
}

@article{jhaver2019does,
  title={Does transparency in moderation really matter? User behavior after content removal explanations on {Reddit}},
  author={Jhaver, Shagun and Bruckman, Amy and Gilbert, Eric},
  journal={Proceedings of the ACM on Human-Computer Interaction},
  year={2019},
  publisher={ACM New York, NY, USA}
}

@article{broniatowski2018weaponized,
  title={Weaponized health communication: {Twitter} bots and {Russian} trolls amplify the vaccine debate},
  author={Broniatowski, David A and Jamison, Amelia M and Qi, SiHua and AlKulaib, Lulwah and Chen, Tao and Benton, Adrian and Quinn, Sandra C and Dredze, Mark},
  journal={American Journal of Public Health},
  year={2018},
  publisher={American Public Health Association}
}

@techreport{ssci2019report,
  author = {{U.S. Senate Select Committee on Intelligence}},
  title = {Report on {Russian} Active Measures Campaigns and Interference in the 2016 {U.S.} Election, Volume 2: {Russia}'s Use of Social Media, with Additional Views},
  year = {2019},
  institution = {United States Senate},
  url = {https://www.intelligence.senate.gov/sites/default/files/documents/Report_Volume2.pdf}
}

@techreport{howard2018ira,
    title       = {The {IRA}, Social Media and Political Polarization in the United States, 2012--2018},
  author      = {Howard, Philip N and Ganesh, Bharath and Liotsiou, Dimitra and Kelly, John and Fran{\c{c}}ois, Camille},
  institution = {Project on Computational Propaganda, University of Oxford},
  year        = {2018}
}

@article{alizadeh2020content,
  title={Content-based features predict social media influence operations},
  author={Alizadeh, Meysam and Shapiro, Jacob N and Buntain, Cody and Tucker, Joshua A},
  journal={Science Advances},
  year={2020}
}

@article{papke1996econometric,
  title={Econometric methods for fractional response variables with an application to 401 (k) plan participation rates},
  author={Papke, Leslie E and Wooldridge, Jeffrey M},
  journal={Journal of Applied Econometrics},
  year={1996},
  publisher={Wiley Online Library}
}

@techreport{kpf2025media,
  author = {{Korea Press Foundation}},
  title = {2025 {Media Users in Korea} (2025 {언론수용자 조사})},
  institution = {Korea Press Foundation},
  year = {2025},
  month = {December},
  address = {Seoul, South Korea},
  url = {https://www.kpf.or.kr/front/research/consumerDetail.do?seq=600224},
  note = {Available at \url{https://www.kpf.or.kr/front/research/consumerDetail.do?seq=600224}}
}

@book{newman2025digital,
  title={Digital news report 2025},
  author={Newman, Nic and Ross Arguedas, Arguedas and Robertson, Craig T and Nielsen, Rasmus Kleis and Fletcher, Richard},
  year={2025},
  publisher={Reuters Institute for the Study of Journalism}
}

@inproceedings{jeong2020identifying,
  title={Identifying and quantifying coordinated manipulation of upvotes and downvotes in {Naver News} comments},
  author={Jeong, Jiwan and Kang, Jeong-han and Moon, Sue},
  booktitle={Proceedings of the International AAAI Conference on Web and Social Media},
  year={2020}
}

@inproceedings{saeed2022trollmagnifier,
  title={Trollmagnifier: Detecting state-sponsored troll accounts on {Reddit}},
  author={Saeed, Mohammad Hammas and Ali, Shiza and Blackburn, Jeremy and De Cristofaro, Emiliano and Zannettou, Savvas and Stringhini, Gianluca},
  booktitle={IEEE symposium on security and privacy},
  year={2022}
}

@inproceedings{kireev2025characterizing,
  title={Characterizing and Detecting Propaganda-Spreading Accounts on {Telegram}},
  author={Kireev, Klim and Mykhno, Yevhen and Troncoso, Carmela and Overdorf, Rebekah},
  booktitle={Proceedings of the 34th USENIX Conference on Security Symposium},
  year={2025}
}

@inproceedings{xiao2025sokML,
author = {Xiao, Madelyne and Mayer, Jonathan},
title = {{SoK}: Machine Learning for Misinformation Detection},
year = {2025},
booktitle = {34th USENIX Security Symposium},
}

@inproceedings{hanley2024specioussites,
  author={Hanley, Hans W. A. and Kumar, Deepak and Durumeric, Zakir},
  booktitle={IEEE Symposium on Security and Privacy}, 
  title={Specious Sites: Tracking the Spread and Sway of Spurious News Stories at Scale}, 
  year={2024}
}

@inproceedings{mirza2023tactics,
  title={Tactics, Threats \& Targets: Modeling Disinformation and its Mitigation},
  author={Mirza, Muhammad Shujaat and Begum, Labeeba and Niu, Liang and Pardo, Sarah and Abouzied, Azza and Papotti, Paolo and P{\"o}pper, Christina},
  booktitle={Network and Distributed System Security Symposium},
  year={2023}
}

\appendix
\section{Troll Dataset Summary}
\label{sec:troll_dataset_summary}

We define the 70 accounts publicly released by the Institute for National Security Strategy in South Korea~\cite{INSS2024ForeignInfluence} as \emph{known trolls}, accounts flagged by our user-level detector as \emph{model-detected trolls}, and all remaining accounts after excluding these two groups as \emph{non-troll users}.

\begin{table}[H]
\centering
\caption{Troll dataset summary by troll category. The total number of comments by known trolls increased from an initial 356,378 to include an additional 3,799 comments identified from the candidate pool. The total number of non-troll users increased from the initial 81 users by adding 4,023,682 additional users.}

\label{tab:troll_user_dataset}
\small
\begin{tabular*}{\columnwidth}{@{\extracolsep{\fill}}ccc}
\toprule
\textbf{Troll Category} & \textbf{Users} & \textbf{Comments}
\\
\midrule
Known Troll      & 70        & 360,177 \\
Model-Detected Troll   & 23,998    & 14,698,055 \\
Non-Troll        & 4,023,763 & 97,600,322 \\
\midrule
\textbf{Total}   & \textbf{4,047,831} & \textbf{112,658,554}
\\
\bottomrule
\end{tabular*}
\end{table}

\begin{table}[h]
\centering
\caption{Hierarchical distribution of classes in the training data ($N=49,745$). The table presents the count and proportion of each class within its respective level. For Levels 2 and 3 (multi-label), only applicable classes are listed. “MNS” refers to \textit{a major neighboring state}.}
\label{tab:train_data_distribution}
\resizebox{\columnwidth}{!}{%
\begin{tabular}{llrc}
\toprule
\multicolumn{1}{c}{\textbf{Level}} & \multicolumn{1}{c}{\textbf{Class}} & \multicolumn{1}{c}{\textbf{Count}} & \multicolumn{1}{c}{\textbf{Proportion}} \\
\midrule
\multirow{2}{*}{Level 1} & Foreign State-suspected & 21,627 & 0.43 \\
 & Not Foreign State-suspected & 28,118 & 0.57 \\
\midrule
\multirow{2}{*}{Level 2} & Other-Condemning & 17,825 & 0.92 \\
 & Other-Praising & 1,608 & 0.08 \\
\midrule
\multirow{10}{*}{Level 3} & Condemning Korea & 12,849 & 0.67 \\
 & Condemning MNS & 3,184 & 0.16 \\ 
 & Condemning Rivals & 2,614 & 0.14 \\
 & Condemning Partners & 112 & 0.01 \\
 & Condemning Others & 349 & 0.02 \\
\cmidrule{2-4}
 & Praising Korea & 89 & 0.05 \\
 & Praising MNS & 1,424 & 0.84 \\
 & Praising Rivals & 86 & 0.05 \\
 & Praising Partners & 76 & 0.05 \\ 
 & Praising Others & 22 & 0.01 \\
\midrule
\multirow{2}{*}{Final} & Troll & 17,751 & 0.36 \\
 & Non-Troll & 31,994 & 0.64 \\
\bottomrule
\end{tabular}%
}
\end{table}

\section{Comparative Analysis of Troll Groups}
\label{sec:comparative_troll}

We analyze the differences between model-detected trolls and non-troll users, showing that model-detected trolls are more similar to known trolls than non-troll users in terms of comment overlap with known trolls and content-level prediction patterns.

\paragraph{Comment Overlap with Known Trolls.} \textit{Exact comment overlap} with known trolls is included as an input feature, but it contributes negligibly in our model (ranked 24th out of 24 by mean absolute SHAP; mean(|SHAP|) = 0), making it unlikely to be a primary driver of detection. We therefore report this analysis to contextualize the prevalence and qualitative nature of duplicated texts, rather than as an independent validation signal. Using the same criteria as the \textit{Exact match overlap with the known trolls} feature (comments with $\ge$10 characters and $\ge$3 tokens), we measure exact-text overlap with comments authored by known trolls. 

\begin{figure}[h]
\centering
\includegraphics[width=0.8\linewidth]{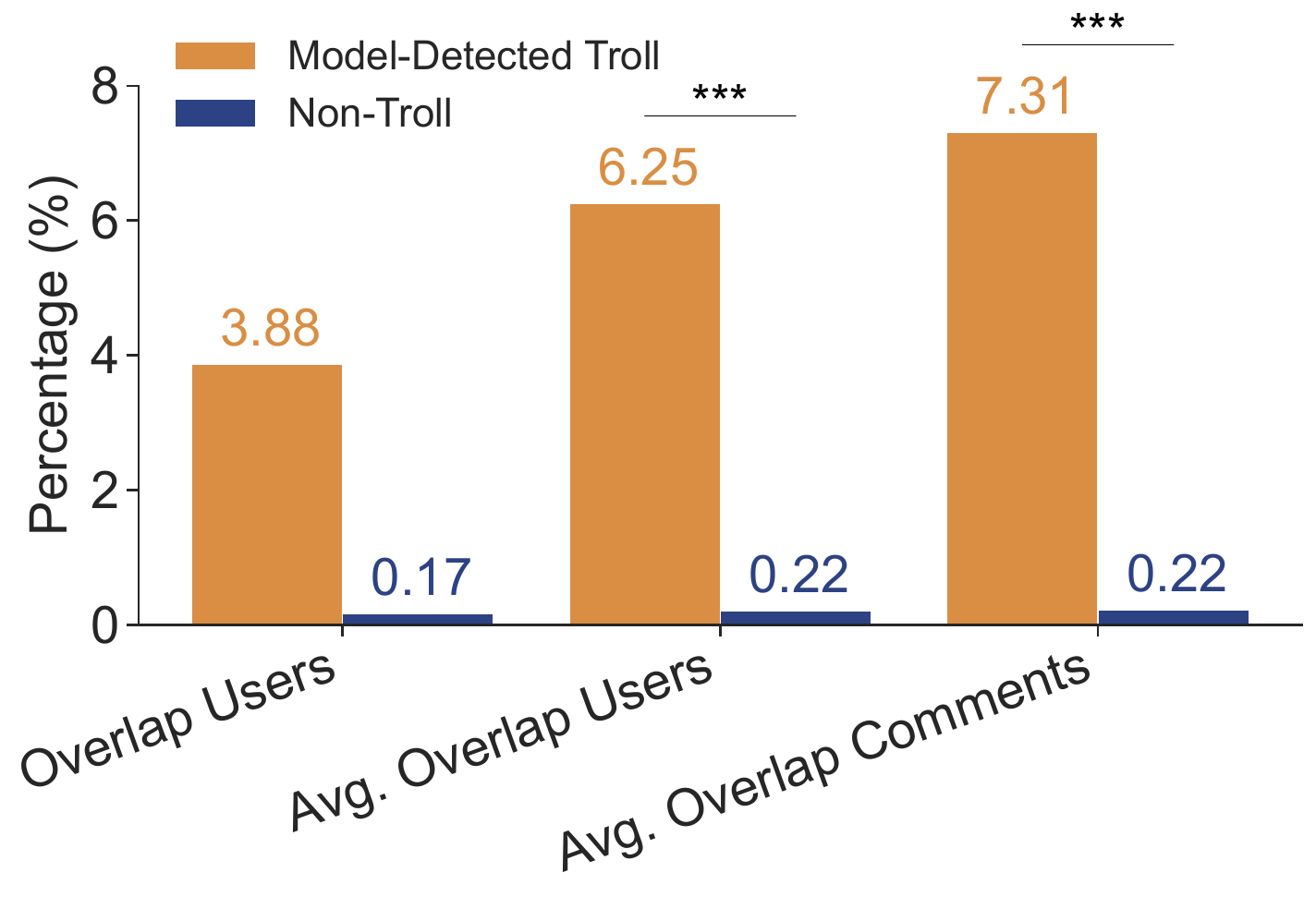}
\caption{Overlap with known trolls by troll group. \textit{Overlap Users} refers to the percentage of users in each group whose comments overlap with those of known trolls. \textit{Average Overlap Users} and \textit{Average Overlap Comments} refer to the average number of known troll users and comments, respectively, that overlap with each user in a group. Asterisks (***) above each class indicate a significant difference in means based on the Mann-Whitney U test ($p<0.001$).}
\label{fig:network_with_known_troll}
\end{figure}

Figure~\ref{fig:network_with_known_troll} shows that model-detected trolls exhibited substantially higher overlap with known trolls: 3.88\% (930 of 23,998) of model-detected troll users shared comments with at least one known troll user, compared to 0.17\% (6,941 of 4 million) of non-troll users. Model-detected trolls also overlapped with more known trolls and shared more matching comments on average than non-troll users (Mann-Whitney U test, $p<0.001$). 

\paragraph{Content-Level Probabilities.} Appendix Table~\ref{tab:mean_class_probability} shows average user-level probabilities, conducted using the same method as in Table~\ref{tab:mean_class_prediction}. Model-detected troll users exhibited significantly higher probabilities across all six classes (\textit{Troll Comments}: \textit{t} = 187.19; \textit{Foreign State-suspected}: \textit{t} = 195.42; \textit{Condemning Korea}: \textit{t} = 184.25; \textit{Condemning Rivals}: \textit{t} = 48.89; \textit{Praising MNS}: \textit{t} = 27.98; \textit{Praising Partners}: \textit{t} = 27.19; all \textit{p} < 0.001).
\begin{table}[H]
\centering
\small
\caption{User-level average percentages computed from content-level probabilities across six classes for model-detected troll and non-troll users. Model-detected trolls consistently exhibit higher probabilities across all six classes that are considered important within the hierarchical detection framework. “MNS” refers to \textit{a major neighboring state}.}
\label{tab:mean_class_probability}
\begin{tabular}{ccc}
\toprule
\textbf{Class} & \textbf{Model-Detected Troll} & \textbf{Non-Troll} \\ \midrule
Troll Comments          & 57.66\% & 18.27\% \\
Foreign State-suspected & 75.20\% & 27.89\% \\
Condemning Korea        & 55.20\% & 17.43\% \\
Condemning Rivals   & 1.48\%  & 0.45\% \\
Praising MNS            & 0.59\%  & 0.20\% \\
Praising Partners   & 0.14\%  & 0.06\% \\
\bottomrule
\end{tabular}
\end{table}

\section{Comparative Follow Relationships}
\label{sec:follow_analysis}
\begin{table*}
\centering
\caption{Follow network analysis of known troll and non-troll users: Comparative metrics}
\label{tab:follow_network_analysis}
\renewcommand{\arraystretch}{1.15}

\newcolumntype{P}[1]{>{\centering\arraybackslash}p{#1}}

\newcolumntype{M}[1]{>{\centering\arraybackslash}m{#1}}

\resizebox{0.95\textwidth}{!}{
\begin{tabular}{
M{1.3cm} 
| M{1.1cm} M{1.1cm}
| M{1.1cm} M{1.1cm}
| M{1.3cm} M{1.3cm}
| M{1.2cm} M{1.2cm}
| M{1.3cm} M{1.3cm} M{1.3cm} M{1.3cm}
}
\toprule
\multirow{2}{*}[-2.0em]{\textbf{Metric}} & 
\multicolumn{2}{c|}{\makecell{\textbf{Followers}}} & 
\multicolumn{2}{c|}{\makecell{\textbf{Followings}}} & 
\multicolumn{2}{c|}{\makecell{\textbf{Reciprocal Follows}}} & 
\multicolumn{2}{c|}{\makecell{\textbf{Follow Ratio}}} & 
\multicolumn{4}{c}{\makecell{\textbf{Known Troll Follow}}} \\
\cmidrule(lr){2-3}\cmidrule(lr){4-5}\cmidrule(lr){6-7}\cmidrule(lr){8-9}\cmidrule(lr){10-13}
& \textbf{Known Troll} & \textbf{Non-Troll} 
& \textbf{Known Troll} & \textbf{Non-Troll} 
& \textbf{Known Troll} & \textbf{Non-Troll} 
& \textbf{Known Troll} & \textbf{Non-Troll} 
& \makecell{\textbf{Known}\\\textbf{Troll}\\\textbf{Followers}}
& 
\makecell{\textbf{Non-}\\\textbf{Troll}\\\textbf{Followers}}
& \makecell{\textbf{Known}\\\textbf{Troll}\\\textbf{Followings}}
& \makecell{\textbf{Non-}\\\textbf{Troll}\\\textbf{Followings}} \\

\midrule
\textbf{Average} & 63.3676 & 23.4464 & 34.1324 & 0.9821 & 2.4559 & 0.0179 & 0.5954 & 0.1166 & 2.2353 & 0 & 2.1324 & 0\\
\textbf{Min}     & 2 & 1 & 0 & 0 & 0 & 0 & 0 & 0 & 0 & 0 & 0 & 0 \\
\textbf{Max}     & 472 & 238 & 499 & 10 & 28 & 1 & 7.9677 & 1.7500 & 11 & 0 & 27 & 0 \\

\bottomrule
\end{tabular}}
\end{table*}

We compare the follow relationships of 70 known troll accounts and 81 non-troll accounts based on their collected followers and followings. The known troll accounts have 4,309 followers and 2,375 followings, whereas the non-troll accounts have 1,313 followers and 55 followings. A detailed comparison is presented in Table~\ref{tab:follow_network_analysis}.

\paragraph{Activity} Known trolls are far more active in forming follow relationships, with on average 2.7 times more followers and 35 times more followings than non-troll accounts. The substantially higher number of followings points to an outward-facing strategy for reaching other users, communities, or potentially influential accounts. In contrast, non-troll users maintain far fewer followings, suggesting that troll following patterns are less consistent with ordinary social ties and more consistent with strategic outreach.

\paragraph{Network} Known trolls form a cohesive network through mutual following. On average, each troll account follows and is followed by two other known trolls, with some mutually connected to up to nine. Reciprocal following occurs 138 times more often among known trolls than among non-troll accounts. Non-troll accounts show no follow relationships with known trolls. This pattern indicates that known troll accounts operate as an internally connected cluster rather than as isolated accounts, a structure that may support coordinated activity or amplification.

\paragraph{Reciprocity} Known trolls also exhibit imbalanced follow relationships. We define the follow ratio as the number of followings divided by the number of followers plus one. This ratio is five times higher for known trolls than for non-troll accounts. This outbound-heavy pattern suggests a strategy of reaching many users, which may help create follow-backs and reciprocal connections.

\section{Robustness to Paraphrasing}
\label{sec:robustness_paraphrase}

\begin{table}[H]
\centering
\caption{Robustness to LLM paraphrasing. BLEU and SBERT cosine similarity are computed between each original test comment and its paraphrase. Macro-F1 is averaged across classes. Performance remains close to the original, and lexical overlap is near zero.}
\label{tab:paraphrase_robustness}
\resizebox{\columnwidth}{!}{%
\begin{tabular}{lcccc}
\toprule
\textbf{Model} & \textbf{Condition} & \textbf{BLEU} & \textbf{SBERT} & \textbf{Macro-F1} \\
\midrule
\textit{Original} & -- & -- & -- & 0.8411 \\
\midrule
GPT-5      & \textit{base} & 0.013 & 0.8148 & 0.8209 \\
GPT-5      & \textit{meta} & 0.010 & 0.8016 & 0.8540 \\
GPT-5-mini & \textit{base} & 0.016 & 0.8274 & 0.8088 \\
GPT-5-mini & \textit{meta} & 0.016 & 0.8120 & 0.8404 \\
\bottomrule
\end{tabular}%
}
\end{table}

A content-level detector may exploit actor- or wording-specific surface signatures rather than the target concepts, which could inflate performance. To test for this, we adapted the LLM-bypassing protocol and paraphrased the 1{,}000 human-annotated test comments using GPT-5 and GPT-5-mini under two conditions~\cite{kireev2025characterizing}. The \textit{base} condition provides only the article title and the comment, requesting a rewrite in the style of a native Korean commenter that preserves meaning, stance, and tone. The \textit{meta-informed} condition additionally provides the annotated moral emotion and the condemning and praising targets, asking the model to preserve these attitudes while rewriting. Because span-level ground truth is not preserved under paraphrasing, we evaluate at the class level and report macro-F1 averaged across classes. The paraphrases substantially alter surface form, with a mean BLEU of 0.0134 relative to the originals, while preserving semantic content, with a mean SBERT cosine similarity of 0.8140. Despite this rewriting, macro-F1 remains comparable to the original 0.8411 across all four settings (0.8088--0.8540; Table~\ref{tab:paraphrase_robustness}). This suggests that the content-level detector does not rely primarily on memorized surface patterns, but instead captures the intended moral-emotional and target-country concepts.

\begin{figure*}
\centering
\includegraphics[width=\textwidth, keepaspectratio]{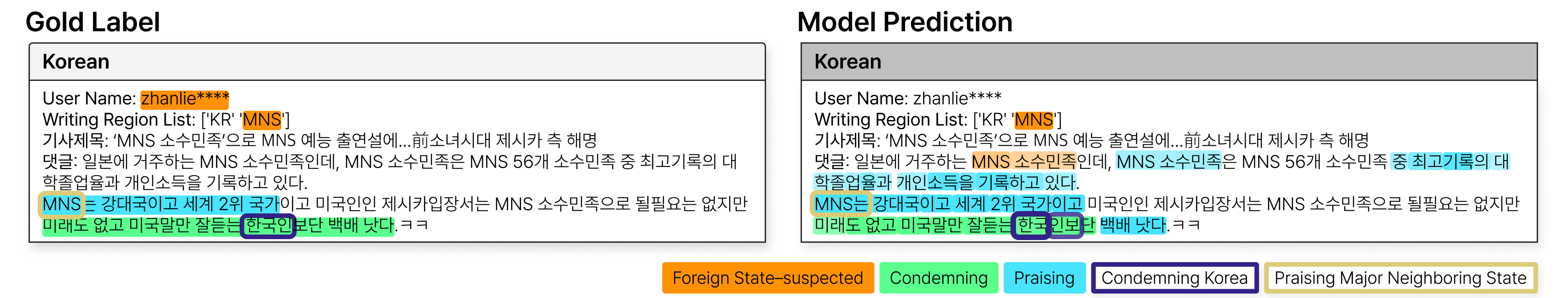}
\caption{Original Korean version of the model output shown in Figure~\ref{fig:rationale_example}, with rationale-related spans highlighted. Each span corresponds to a predicted label: \textit{Foreign state--suspected} (filled orange), \textit{Other-condemning} (filled green), \textit{Other-praising} (filled sky blue), \textit{Condemning Korea} (outlined navy), and \textit{Praising MNS} (outlined gold). Span opacity indicates the predicted probability for that label. ``MNS'' refers to \textit{a major neighboring state}.}
\label{fig:rationale_example_kor}
\end{figure*}

\begin{figure*}[t!]
\centering
\includegraphics[width=0.99\textwidth]{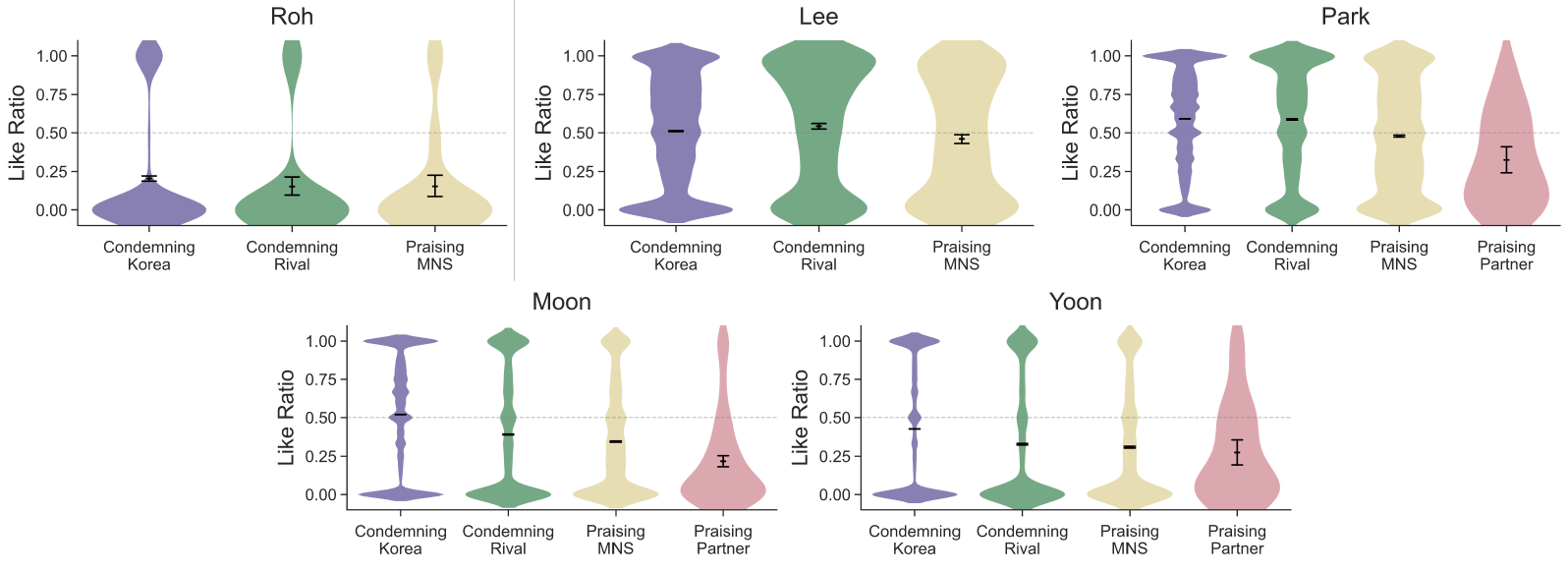}
\caption{Like ratio by rhetoric across Korean administrations. Across five administrations, \textit{Condemning Korea} yields higher like ratios than praising rhetoric, ranking first in four of five administrations; the only exception is the Lee administration, where \textit{Condemning Rival} ranks first. \textit{Praising MNS} and \textit{Praising Partner} are consistently lower than condemning rhetoric across regimes. \textit{Praising Partner} is not shown for the Roh and Lee administrations due to the absence of data. Error bars indicate mean $\pm$ 95\% CI; the dashed line marks 0.5 (neutral: likes equal dislikes).}
\label{fig:like_ratio_across_administration}
\end{figure*}


\begin{table*}[bh!]
\centering
\scriptsize 
\caption{Top 10 most frequent target spans in high-visibility \textit{Condemning Korea} comments (Like Ratio = 1.0) by administration. Politically salient figures, including elected leaders, candidates, and other politically connected individuals, are shown in bold with colored backgrounds. Blue indicates left-leaning figures, and red indicates right-leaning figures.}
\label{tab:target_span_full_list_updated}
\renewcommand{\arraystretch}{1.8} 

\newcolumntype{Y}{>{\centering\arraybackslash}X}

\begin{tabularx}{\textwidth}{YYYYY}
\toprule
\begin{tabular}[t]{@{}c@{}}\textbf{Roh} \\ \scriptsize (Apr, 2006 -- Feb, 2008)\end{tabular} & 
\begin{tabular}[t]{@{}c@{}}\textbf{Lee} \\ \scriptsize (Feb, 2008 -- Feb, 2013)\end{tabular} & 
\begin{tabular}[t]{@{}c@{}}\textbf{Park} \\ \scriptsize (Feb, 2013 -- May, 2017)\end{tabular} & 
\begin{tabular}[t]{@{}c@{}}\textbf{Moon} \\ \scriptsize (May, 2017 -- May, 2022)\end{tabular} & 
\begin{tabular}[t]{@{}c@{}}\textbf{Yoon} \\ \scriptsize (May, 2022 -- Mar, 2025)\end{tabular} \\ \midrule

\begin{tabular}[c]{@{}c@{}}Korea \\ 한국\end{tabular} & 
\begin{tabular}[c]{@{}c@{}}Korea \\ 한국\end{tabular} & 
\begin{tabular}[c]{@{}c@{}}Korea \\ 한국\end{tabular} & 
\cellcolor[HTML]{ADD8E6}\begin{tabular}[c]{@{}c@{}}\textbf{Moon Jae-in} \\ \textbf{문재인}\end{tabular} & 
\cellcolor[HTML]{ADD8E6}\begin{tabular}[c]{@{}c@{}}\textbf{Lee Jae-myung} \\ \textbf{이재명}\end{tabular} \\

\cellcolor[HTML]{ADD8E6}\begin{tabular}[c]{@{}c@{}}\textbf{Roh Moo-hyun} \\ \textbf{노무현}\end{tabular} & 
\begin{tabular}[c]{@{}c@{}}Rep. of Korea \\ 대한민국\end{tabular} & 
\cellcolor[HTML]{FFD1DC}\begin{tabular}[c]{@{}c@{}}\textbf{Park Geun-hye} \\ \textbf{박근혜}\end{tabular} &
 
\cellcolor[HTML]{ADD8E6}\begin{tabular}[c]{@{}c@{}}\textbf{Mun Jae-ang (Derogatory)} \\ \textbf{문재앙}\end{tabular} & 
\cellcolor[HTML]{FFD1DC}\begin{tabular}[c]{@{}c@{}}\textbf{Yoon (Hanja)} \\ \textbf{尹}\end{tabular} \\

\begin{tabular}[c]{@{}c@{}}Our country \\ 우리나라\end{tabular} & 
\begin{tabular}[c]{@{}c@{}}Our country \\ 우리나라\end{tabular} & 
\begin{tabular}[c]{@{}c@{}}Rep. of Korea \\ 대한민국\end{tabular} & 
\cellcolor[HTML]{ADD8E6}\begin{tabular}[c]{@{}c@{}}\textbf{Pres. Moon (Hanja)} \\ \textbf{文대통령}\end{tabular} & 
\cellcolor[HTML]{FFD1DC}\begin{tabular}[c]{@{}c@{}}\textbf{Yoon Suk-yeol} \\ \textbf{윤석열}\end{tabular} \\

\begin{tabular}[c]{@{}c@{}}Rep. of Korea \\ 대한민국\end{tabular} & 
\begin{tabular}[c]{@{}c@{}}LG Electronics \\ LG\end{tabular} & 
\cellcolor[HTML]{FFD1DC}\begin{tabular}[c]{@{}c@{}}\textbf{Park (Hanja)} \\ \textbf{朴대}\end{tabular} & 
\cellcolor[HTML]{ADD8E6}\begin{tabular}[c]{@{}c@{}}\textbf{Cho Kuk} \\ \textbf{조국}\end{tabular} & 
\begin{tabular}[c]{@{}c@{}}Democratic Party \\ 민주당\end{tabular} \\

\cellcolor[HTML]{ADD8E6}\begin{tabular}[c]{@{}c@{}}\textbf{Kim Dae-jung} \\ \textbf{김대중}\end{tabular} & 
\cellcolor[HTML]{ADD8E6}\begin{tabular}[c]{@{}c@{}}\textbf{Moon Jae-in} \\ \textbf{문재인}\end{tabular} & 
\cellcolor[HTML]{FFD1DC}\begin{tabular}[c]{@{}c@{}}\textbf{Choi Soon-sil} \\ \textbf{최순실}\end{tabular} & 
\begin{tabular}[c]{@{}c@{}}Democratic Party \\ 민주당\end{tabular} & 
\cellcolor[HTML]{FFD1DC}\begin{tabular}[c]{@{}c@{}}\textbf{Pres. Yoon} \\ \textbf{윤 대통령}\end{tabular} \\

\begin{tabular}[c]{@{}c@{}}Samsung \\ 삼성\end{tabular} & 
\begin{tabular}[c]{@{}c@{}}Korea (+ Particle) \\ 한국은\end{tabular} & 
\cellcolor[HTML]{FFD1DC}\begin{tabular}[c]{@{}c@{}}\textbf{Pres. Park (Hanja)} \\ \textbf{朴대통령}\end{tabular} & 
\cellcolor[HTML]{ADD8E6}\begin{tabular}[c]{@{}c@{}}\textbf{Moon (Hanja)} \\ \textbf{文}\end{tabular} & 
\cellcolor[HTML]{FFD1DC}\begin{tabular}[c]{@{}c@{}}\textbf{Kim Keon-hee} \\ \textbf{김건희}\end{tabular} \\

\cellcolor[HTML]{FFD1DC}\begin{tabular}[c]{@{}c@{}}\textbf{Lee Myung-bak} \\ \textbf{이명박}\end{tabular} & 
\begin{tabular}[c]{@{}c@{}}Jeolla-do \\ 전라도\end{tabular} & 
\begin{tabular}[c]{@{}c@{}}Hell Joseon \\ 헬조선\end{tabular} & 
\begin{tabular}[c]{@{}c@{}}Korea \\ 한국\end{tabular} & 
\cellcolor[HTML]{ADD8E6}\begin{tabular}[c]{@{}c@{}}\textbf{Moon Jae-in} \\ \textbf{문재인}\end{tabular} \\

\begin{tabular}[c]{@{}c@{}}Chosun Ilbo \\ 조선일보\end{tabular} & 
\cellcolor[HTML]{FFD1DC}\begin{tabular}[c]{@{}c@{}}\textbf{Lee Myung-bak} \\ \textbf{이명박}\end{tabular} & 
\begin{tabular}[c]{@{}c@{}}Korea (+ Particle) \\ 한국은\end{tabular} & 
\cellcolor[HTML]{FFD1DC}\begin{tabular}[c]{@{}c@{}}\textbf{Yoon Suk-yeol} \\ \textbf{윤석열}\end{tabular} & 
\begin{tabular}[c]{@{}c@{}}Rep. of Korea \\ 대한민국\end{tabular} \\

\begin{tabular}[c]{@{}c@{}}Korea (+ Particle) \\ 한국은\end{tabular} & 
\begin{tabular}[c]{@{}c@{}}Korea (short) \\ 한\end{tabular} & 
\begin{tabular}[c]{@{}c@{}}Sewol Ferry \\ 세월호\end{tabular} & 
\begin{tabular}[c]{@{}c@{}}Rep. of Korea \\ 대한민국\end{tabular} & 
\cellcolor[HTML]{FFD1DC}\begin{tabular}[c]{@{}c@{}}\textbf{Han Dong-hoon} \\ \textbf{한동훈}\end{tabular} \\

\begin{tabular}[c]{@{}c@{}}South Korea \\ 남한\end{tabular} & 
\begin{tabular}[c]{@{}c@{}}Journalist \\ 기자\end{tabular} & 
\begin{tabular}[c]{@{}c@{}}Our country \\ 우리나라\end{tabular} & 
\cellcolor[HTML]{ADD8E6}\begin{tabular}[c]{@{}c@{}}\textbf{Lee Jae-myung} \\ \textbf{이재명}\end{tabular} & 
\cellcolor[HTML]{FFD1DC}\begin{tabular}[c]{@{}c@{}}\textbf{Yoon} \\ \textbf{윤}\end{tabular} \\ \bottomrule

\multicolumn{5}{l}{\textit{* Note: Hanja denotes Chinese characters used in Korean; Particle refers to Korean grammatical markers (e.g., -eun/-neun).}} \\
\end{tabularx}
\end{table*}

\end{document}